\documentclass[11pt, a4paper]{article}
\usepackage{amssymb,amsmath,latexsym,graphicx, bm, mathrsfs}
\usepackage[hidelinks,citecolor=blue,urlcolor=blue,colorlinks=true,bookmarks=false,hypertexnames=true]{hyperref} 
\usepackage{natbib} 
\setcitestyle{numbers,square} 
\usepackage{graphics,graphicx, cancel}
\usepackage[dvipsnames]{xcolor}
\usepackage{graphicx}
\graphicspath{ {./images/} }
\usepackage[british]{datetime2}
\usepackage[page,toc,titletoc,title]{appendix}
\usepackage{tocloft}
\usepackage{blindtext}
\usepackage{slashed}
\usepackage{bbm}
\usepackage{dsfont}
\usepackage{dirtytalk}

\usepackage{comment}
\usepackage{caption}
\usepackage{enumitem}
\usepackage{mathrsfs}
\usepackage{alphalph} 
\usepackage{changepage}
\newcounter{mainthm}
 \usepackage{float}

\newcounter{subthm}[mainthm]

\newtheorem{subtheoreminner}{Theorem}[subthm]

\begin{document}

\title{Hawking--Page phase transition for pure Lovelock black holes
\vspace{-0.25cm} 
\author{Nitesh K. Dubey$ ^{a,b}$, Sanved Kolekar$^{a,b}$ \\ \vspace{-0.5cm} \\ 
\textit{$ ^a $Indian Institute of Astrophysics} \\ 
\textit{Block 2, 100 Feet Road, Koramangala,} 
\textit{Bengaluru 560034, India.} \\ 
\textit{$ ^b $Pondicherry University} \\ 
\textit{R.V.Nagar, Kalapet, Puducherry-605014, India}\\
\texttt{\small Email: \href{mailto:nitesh.dubey@iiap.res.in}{nitesh.dubey@iiap.res.in}, 
\href{mailto:sanved.kolekar@iiap.res.in}{sanved.kolekar@iiap.res.in}}}}

\maketitle

\abstract{We investigate the thermodynamic properties of static, spherically symmetric Anti-de Sitter (AdS) black holes, focusing on the interplay between characteristic temperatures, as well as on the universality of Ruppeiner scalar curvature at the Hawking-Page (HP) phase transition. In particular, we study the relation between the minimum temperature and the HP phase transition temperature for static, spherically symmetric AdS black holes in pure Lovelock gravity. For the electromagnetically neutral case in Einstein gravity, the minimum temperature in $(d+1)$ dimensions coincides with the HP transition temperature in $d$ dimensions, while in higher pure Lovelock theories this relation is modified by a dimension- and order-dependent factor, reducing to the Einstein result in appropriate limits. For charged AdS black holes, in the grand canonical ensemble, in general relativity, the two temperatures differ by a simple dimension-dependent factor, whereas no universal relation persists in higher curvature pure Lovelock theories. We further analyze the normalized Ruppeiner scalar curvature at the HP transition and show that it is a universal constant depending only on the spacetime dimension for electromagnetically neutral black holes in pure Lovelock theories. The normalized scalar curvature remains a constant, under appropriate conditions, even for the charged static spherically symmetric black holes in the grand canonical ensemble for the Einstein theory case, whereas in general pure Lovelock theories it depends on thermodynamic parameters such as pressure and electrostatic potential, asymptotically approaching a constant in the large-pressure or simultaneous large-potential and large-pressure limits.}

\pagebreak

\section{Introduction}

Black hole thermodynamics provides a well-controlled framework for probing the interplay between gravity, quantum theory, and statistical mechanics. In anti-de Sitter (AdS) spacetime, the Hawking temperature of Schwarzschild–AdS black holes exhibits a minimum as a function of the horizon radius, below which there is no equilibrium AdS black-hole solution in the fixed temperature ensemble \cite{Belhaj:2020mdr}. Above this minimum temperature, two branches of solutions appear: a small black hole branch with negative specific heat (thermodynamically unstable) and a large black hole branch with positive specific heat. The large black hole branch remains metastable below a temperature called the Hawking--Page phase transition temperature. The Hawking–Page phase transition \cite{Hawking:1982dh} describes a first-order transition between the thermal radiation phase in an AdS spacetime and a stable large black hole configuration. This transition encodes the thermodynamic preference of geometries in a canonical ensemble and acquires a deeper interpretation within the gauge/gravity duality \cite{Witten:1998zw, Aharony:1999ti, Landsteiner:1999gb}, where it is mapped to a confinement–deconfinement transition in the dual large-$N$ gauge theory. Additional thermodynamic and dynamical instabilities may arise depending on the ensemble and the underlying gravitational theory. \cite{ Belhaj:2020mdr, Gregory:1993vy, PhysRevD.102.024085, Chamblin:1999tk, Myung:2013uka, 731x-38lt}.

A natural direction for exploring quantum gravitational effects is to examine modifications of General Relativity involving higher-curvature and higher-derivative corrections. Such terms generically arise in the low-energy effective action of string theory \cite{Zwiebach:1985uq, Gross:1986iv} and can significantly alter both the geometric and thermodynamic properties of black hole solutions. In higher dimensions, these corrections are systematically incorporated through Lovelock gravity \cite{Lovelock:1971yv}, the most general local diffeomorphism-invariant metric theory whose Lagrangian is constructed solely from the metric and the Riemann tensor and whose metric field equations are second order. Black objects in higher dimensions, such as black strings and black branes, are often constructed by extending lower-dimensional solutions along additional flat directions. While this procedure is not universally valid in arbitrary higher-curvature theories, it has been shown to hold for a class of Lovelock theories admitting (locally) unique constant-curvature vacua \cite{Kastor:2006vw}. This observation has led to the formulation of pure Lovelock gravity, in which only a single Lovelock term of order $n$ (along with a cosmological constant) is retained in the action.

Pure Lovelock gravity constitutes a particularly clean and minimal extension of Einstein gravity to higher dimensions. Unlike generic Lovelock theories, where multiple curvature invariants contribute simultaneously and lead to an intricate solution space, the pure theory isolates the dynamical role of a single higher-curvature term. Despite this simplification, it preserves key structural features, including second-order field equations, absence of ghost-like excitations around suitable backgrounds, and a natural generalization of geometric properties familiar from General Relativity \cite{Dadhich:2012ma, Dadhich:2012zq, Gannouji:2013eka}. Importantly, pure Lovelock models are not merely effective corrections emerging from an underlying theory; rather, they define consistent classical gravitational theories in their own right \cite{Gannouji:2013eka}. This makes them particularly suitable for isolating and studying the qualitative impact of higher-curvature interactions on black hole physics. An additional motivation for studying pure Lovelock black holes arises from the remarkable universality exhibited by their thermodynamic properties \cite{Kolekar:2010dm, Kolekar:2011gw, Kolekar:2011gg, Kolekar:2012tq, Padmanabhan:2013xyr}. In particular, several thermodynamic quantities exhibit scaling properties that can be expressed in terms of an effective dimension determined by the Lovelock order $n$ \cite{Dadhich:2013bya}. In the static, spherically symmetric sector, the corresponding black hole solutions remain analytically tractable and closely parallel those of Einstein gravity, facilitating a systematic comparison. Pure Lovelock gravity also provides a useful setting for examining the microscopic origin of black hole entropy. In General Relativity, different approaches to entropy, such as geometrical, thermodynamic, and semiclassical, converge to the Bekenstein area law. In Lovelock theories, however, these notions need not coincide trivially, and the higher-curvature contributions can help distinguish between competing interpretations and reveal the geometric origin of black hole entropy more clearly \cite{Kolekar:2011bb}.

Studies of Hawking-Page transition and related thermodynamic phenomena provide an important background for investigating phase structure of black holes in pure Lovelock gravity. Ref.~\cite{Estrada:2019cig} demonstrated that charged asymptotically AdS black holes in Pure Lovelock gravity exhibit behavior analogous to that of a Van der Waals fluid. It was found that the system undergoes a first-order phase transition between small and large stable black holes, reminiscent of the liquid-gas phase transition in a conventional Van der Waals fluid. Ref.~\cite{Khuzani} further investigated phase transitions in charged pure Lovelock black holes, demonstrating correspondence between thermodynamic-curvature singularities and heat-capacity divergences, studying universal critical scaling, and using Ehrenfest relations to characterize second-order transitions; in extended phase space, the Gibbs free energy exhibits the swallowtail structure associated with first-order small/large black-hole transitions.

Related results in other gravitational settings provide further context. Ref.~\cite{MYUNG2005297} studied the thermodynamics of Lovelock black holes and found that Schwarzschild-AdS black holes in even dimensions exhibit the conventional Hawking-Page transition, whereas Chern-Simons black holes in odd dimensions need not. However, in three dimensions, \cite{MYUNG2005297} found a possible second-order transition between the non-rotating BTZ black hole and the thermal AdS space. Ref.~\cite{PhysRevD.90.064028} studied thermalon-mediated phase transitions between different branches of Lovelock gravity, characterized by distinct effective cosmological constants, and showed that above a critical temperature such transitions can trigger a generalised version of the Hawking-Page transition. Other approaches to black-hole phase transitions have also been explored. For example, Ref.~\cite{Majumdar_2007} discussed a statistical-mechanical generalization using isolated horizons and micro canonical entropy calculated in Loop Quantum Gravity, while Ref.~\cite{Spallucci2013Maxwell} proposed a Maxwell equal-area construction for Schwarzschild-AdS black holes that eliminates the negative-heat-capacity branch and modifies the conventional Hawking-Page transition, yielding a transition temperature \(T_*\) below the standard Hawking-Page temperature, with pure radiation favored for \(T<T_*\) and stable black holes for \(T>T_*\), without an intermediate unstable black-hole phase. More broadly, thermodynamic properties, critical phenomena, and phase transitions of Lovelock black holes have been widely studied in the literature, including extended phase space thermodynamics and holographic aspects~\cite{Kastor:2010gq,Cai:2013qga,Frassino:2014pha}. The dynamics of pure Lovelock gravity itself also exhibits interesting dimensional and background dependence, Ref.~\cite{PhysRevD.93.064009} analyzed the local symmetries and propagating degrees of freedom and showed that the number of physical degrees of freedom depends nontrivially on the background geometry, varying from zero to the maximal value allowed by the theory.  \cite{Wang:2022ska} presents the principle of corresponding states in black hole thermodynamics, and numerically confirms the probe of thermodynamic geometry of the RN–AdS black hole through non-local observables in the dual field theory. 

A particularly relevant development for the present work is the dimensional duality identified in AdS black-hole thermodynamics. Ref.~\cite{PhysRevD.102.104011} uncovered an intriguing dimensional correspondence in terms of a novel dual relation in  AdS black hole thermodynamics in Einstein's gravity by showing that the Hawking-Page transition temperature of a Schwarzschild AdS black hole in $d$ dimensions coincides with the minimum temperature in $(d+1)$ dimensions. ~\cite{PhysRevD.102.104011} also investigated associated aspects of black hole phase structure and thermodynamic geometry in AdS spacetimes in general relativity, particularly through the Ruppeiner framework, where geometric quantities were shown to encode critical behavior and phase transitions. Such a dual relation was subsequently studied in 
Gauss–Bonnet gravity and massive gravity in Refs.~\cite{LIANG2024116673,PhysRevD.104.104049}, suggesting the dual relation may extend beyond Einstein's gravity. 
These developments motivate the study of analogous thermodynamic structures in pure Lovelock gravity. 

In particular, (i) the possible existence of a dimensional thermodynamic duality relation, analogous to those observed for Einstein and Gauss-Bonnet black holes, has not been systematically investigated in the pure Lovelock framework. Likewise, (ii) the role of thermodynamic geometry and the behavior of the Ruppeiner scalar curvature near the Hawking--Page phase transition remain largely unclear in these theories. Since pure Lovelock gravity exhibits several remarkable properties, such as the kinematic nature of gravity in critical odd dimensions and a universal characterization of static black hole solutions, it provides a particularly natural arena to test the robustness of such thermodynamic relations. Moreover, unlike generic higher-curvature theories, pure Lovelock gravity isolates a single Lovelock order, allowing one to identify genuine higher-curvature effects without interference from lower-order contributions. Understanding the thermodynamic phase structure and geometric aspects of pure Lovelock black holes may therefore provide further insight into the universality of AdS black hole thermodynamics and the microscopic interpretation of gravitational phase transitions.
The primary objectives of this paper are therefore to investigate these two aspects of the thermodynamics of pure Lovelock AdS black holes and to determine whether the corresponding relations exhibit a universal character across Lovelock order and spacetime dimension.

The extension to pure Lovelock gravity is not a straightforward generalization of the known Einstein or Gauss--Bonnet results. In particular, the duality relation at the Hawking--Page transition is governed by two distinct conditions, corresponding to the global phase transition and the minimum temperature. While the Hawking--Page condition is determined by the competition between the free energies of the black-hole and thermal-AdS phases, the minimum-temperature condition is fixed by the extremization of the black-hole temperature with respect to the horizon radius. In pure Lovelock gravity, these two conditions involve different $n$-dependent powers of the horizon radius, thereby modifying the relation between the two characteristic temperatures. Consequently, the dimensional relation between the Hawking--Page temperature and the minimum temperature is no longer expected to follow directly from the corresponding Einstein or Gauss--Bonnet results. This difference is particularly important because it provides a natural setting to investigate whether a generalized thermodynamic duality survives in higher-order Lovelock theories and, if so, how it depends on the Lovelock order and spacetime dimension.

In this work, we investigate the thermodynamics of static, spherically symmetric AdS black hole solutions in pure Lovelock gravity within the fixed electrostatic potential ensemble, commonly referred to as the grand canonical ensemble \cite{PhysRevD.60.064018, Myung:2008dma, PhysRevD.93.124021, Mo:2015uwa, PhysRevD.102.124015}. This ensemble provides a natural setting to investigate the thermodynamic phase structure of charged black holes. In particular, Ref. \cite{Huang:2021iyf} found that, in the grand canonical ensemble with fixed potential, a black hole undergoes a first-order phase transition as the temperature increases when the potential is low, whereas no phase transition occurs when the electrostatic potential is high.  The interplay between electric charge, higher-curvature effects, and the AdS length scale gives rise to nontrivial thermodynamic behavior, including the existence of multiple black hole branches, Hawking--Page phase transition\cite{Kubiznak:2012wp, PhysRevD.111.084072}, and potential instabilities. We analyze these features in detail and elucidate how the presence of a single higher-curvature term modifies the familiar thermodynamic landscape of charged AdS black holes. 

We begin the Sec.~\ref{subsec:geom_thermo_setup} with a brief review of the black hole solutions and thermodynamic quantities in general relativity and pure Lovelock gravity. In Sec.\ref{sec:temp_relation}, we first consider chargeless spherically symmetric pure Lovelock AdS black holes and show that the minimum temperature in $(d+1)$ dimensions differs from the Hawking-Page transition temperature in $d$ dimensions just by a spacetime dimension-dependent factor. In the case of Einstein gravity, the extra factor becomes unity, and the minimum temperature in $(d+1)$ dimensions coincides exactly with the Hawking--Page transition temperature in $d$ dimensions. We further extend our analysis to charged black holes in Sec.\ref{sec:curvature}, where the presence of an additional scale leads to qualitatively different behavior. In particular, while a simple dimension-dependent dual relation between the minimum temperature and the Hawking--Page transition temperature persists in general relativity, no such universality survives in higher-order pure Lovelock black holes.

To gain further insight into the geometric aspects of the Hawking--Page (HP) phase transition, we also study the Ruppeiner scalar curvature evaluated at the HP phase transition point. This provides a complementary probe of the phase structure, enabling us to identify universal features that are not immediately evident from thermodynamic quantities such as the Gibbs free energy, temperature, and pressure. Our analysis in Section~\ref{sec:curvature} reveals that while Einstein gravity exhibits a high degree of universality, a constant normalized scalar curvature at the HP phase transition, the higher order pure Lovelock introduces parameter dependence that enriches the structure, albeit with simplifications emerging in certain limits. Most importantly, we show that the HP phase transition doesn't take place above a certain large electrostatic potential for a fixed set of all other parameters. However, if one increases both the pressure as well as the electrostatic potential simultaneously, one gets a constant normalized scalar curvature at the HP phase transition even for higher order pure Lovelock black holes. We conclude in section~\ref{sec:conclusion} with a discussion of the implications and possible future directions. We adopt mostly-plus signature and the natural units where $\hbar = c=G_N=1$.

\section{Charged black holes in pure Lovelock theories}\label{subsec:geom_thermo_setup}

To investigate how higher-curvature terms modify the physics of charged black holes, we consider $d$-dimensional pure Lovelock gravity, in which the
gravitational dynamics are governed by a single $n$-th order Lovelock term,
supplemented by a negative cosmological constant\cite{Lovelock1971,CaiOhta2006}. The theory is minimally
coupled to a Maxwell field, thereby providing a natural framework for
exploring the interplay between higher-curvature gravitational effects,
electromagnetic charge, and the asymptotically AdS geometry\cite{EstradaAros2020}.In particular,
the $n$-th order Lovelock term encodes the higher-curvature contribution to
the gravitational dynamics, while the negative cosmological constant
introduces a cosmological length scale $\ell$. In higher-order pure Lovelock
gravity, the effective asymptotic AdS curvature radius is determined jointly
by $\ell$ and the Lovelock coupling. The corresponding bulk action is
\begin{equation}
I[g,A]
=
\frac{1}{16\pi}
\int_{\mathcal{M}} d^d x\,\sqrt{-g}\,
\left[
\alpha_n \mathcal{L}_n
-2\Lambda
-F_{\mu\nu}F^{\mu\nu}
\right],
\label{eq:action}
\end{equation}
where
\begin{equation}
F_{\mu\nu}
=
\partial_\mu A_\nu-\partial_\nu A_\mu,
\qquad
\Lambda
=
-\frac{(d-1)(d-2)}{2\ell^2},
\label{eq:lambda}
\end{equation}
and $\mathcal{L}_n$ is the $n$th-order Lovelock density\cite{Lovelock1971}.,
\begin{equation}
\mathcal{L}_n
=
\frac{1}{2^n}
\delta^{\mu_1\nu_1\cdots\mu_n\nu_n}
_{\alpha_1\beta_1\cdots\alpha_n\beta_n}
\prod_{p=1}^{n}
R^{\alpha_p\beta_p}{}_{\mu_p\nu_p}.
\label{eq:lovelock_density}
\end{equation}

Variation of the action with respect to the metric and the gauge
potential gives, respectively,
\begin{equation}
\alpha_n \mathcal{G}^{(n)}_{\mu\nu}
+\Lambda g_{\mu\nu}
=
8\pi T_{\mu\nu}^{(F)},
\label{eq:lovelock_eom}
\end{equation}
and
\begin{equation}
\nabla_\mu F^{\mu\nu}=0,
\label{eq:maxwell_eom}
\end{equation}
where
\begin{equation}
T_{\mu\nu}^{(F)}
=
\frac{1}{4\pi}
\left(
F_{\mu\rho}F_{\nu}{}^{\rho}
-\frac{1}{4}g_{\mu\nu}
F_{\rho\sigma}F^{\rho\sigma}
\right),
\label{eq:maxwell_stress}
\end{equation}
and the $n$th-order Lovelock tensor is
\begin{equation}
\mathcal{G}^{(n)\mu}{}_{\nu}
=
-\frac{1}{2^{n+1}}
\delta^{\mu\mu_1\nu_1\cdots\mu_n\nu_n}
_{\nu\alpha_1\beta_1\cdots\alpha_n\beta_n}
\prod_{p=1}^{n}
R^{\alpha_p\beta_p}{}_{\mu_p\nu_p}.
\label{eq:lovelock_tensor}
\end{equation}

The above field equations provide the general dynamical framework of the theory. We now restrict to the static, spherically symmetric sector relevant for the charged black-hole solutions considered below. With this symmetry, the Maxwell equation can be integrated to determine the electric field in terms of the conserved charge. We begin by specifying the spacetime ansatz and thermodynamic conventions that will be used in subsequent sections to study equilibrium black holes that are static and spherically symmetric in $d$-dimensional asymptotically anti-de Sitter (AdS) spacetime. In this sector, one may employ the standard Schwarzschild-type gauge, in which the geometry is completely characterized by a single function $f(r)$\cite{Zegers2005,CaiOhta2006}. This parametrization is particularly convenient because the outer horizon location $r_+$, defined by the largest positive root of $f(r_+)=0$, provides a natural convenient parameter for the thermodynamics. The asymptotic branch is fixed by requiring that, for large $r$, the metric
approaches an AdS geometry, $f(r)\sim 1+r^2/L_{\rm eff}^2,$ as $r\to\infty,$
where $L_{\rm eff}$ denotes the effective AdS curvature radius. In pure
Lovelock gravity, $L_{\rm eff}$ is in general determined by both the
cosmological scale $\ell$ and the Lovelock coupling\cite{DadhichPonsPrabhu2013,CamanhoEdelstein2013}.

To solve the coupled system~\eqref{eq:lovelock_eom}--\eqref{eq:maxwell_eom},
we look for static, spherically symmetric solutions  ansatz of the form\cite{PhysRevD.74.064001, Khuzani:2022lqd}
\begin{equation}\label{eq:purLovelockMetric}
ds^2 = - f(r)\, dt^2 + f(r)^{-1} dr^2 + r^2 d\Sigma_{d-2}^2.
\end{equation}
The angular part of the metric is taken to be the round line element $d\Sigma_{d-2}^2$ on the unit $(d-2)$-sphere. We can introduce its volume as
\begin{equation}
\Sigma_{d-2}=\frac{2\pi^{(d-1)/2}}{\Gamma\!\left(\frac{d-1}{2}\right)},
\end{equation}
so that extensive quantities (mass, charge, entropy) carry their standard normalization in an arbitrary dimension. Substituting the ansatz~\eqref{eq:purLovelockMetric},
together with the static, spherically symmetric gauge potential
$A_\mu = \left(\Phi(r),0,\dots,0\right)$ solving
Eq.~\eqref{eq:maxwell_eom}, into the field
equation~\eqref{eq:lovelock_eom} reduces the gravitational field equations
to a single first-order differential equation, which can be integrated once
to obtain an algebraic relation for the combination $(1-f(r))/r^2$.
Equivalently, the integrated equation may be written as
\begin{equation}
\hat{\alpha}_n
\left(\frac{1-f(r)}{r^2}\right)^n
=
\frac{16\pi M}
{\Sigma_{d-2}(d-2)\,r^{d-1}}
-
\frac{32\pi^2 Q^2}
{\Sigma_{d-2}^2(d-3)(d-2)\,r^{2(d-2)}}
-
\frac{1}{\ell^2},
\label{eq:integrated_lovelock}
\end{equation}
for $d>3$. Solving this algebraic relation for $f(r)$ yields
\begin{equation} \label{eq:f(r)}
f(r) = 1 \pm r^2 \left(
\frac{16\pi M}{\Sigma_{d-2} \hat{\alpha}_n (d-2)\, r^{d-1}}
- \frac{32\pi^2 Q^2}{\Sigma_{d-2}^2 \hat{\alpha}_n (d-3)(d-2)\, r^{2(d-2)}}
- \frac{1}{\hat{\alpha}_n \ell^2}
\right)^{1/n}.
\end{equation}
Here, $M$ and $Q$ denote the mass and electric-charge parameters,
respectively. Since Eq.~\eqref{eq:integrated_lovelock} is an $n$th-order
algebraic relation in $(1-f)/r^2$, taking its $n$th root requires a choice of
branch. The physically relevant branch is chosen by demanding that the
metric be real, possess the desired event horizon, and approach the
appropriate AdS vacuum at large radius. We therefore define the physical
branch by the asymptotic condition
\begin{equation}
\frac{1-f(r)}{r^2}
\longrightarrow
-\frac{1}{L_{\rm eff}^2},
\qquad r\to\infty,
\label{eq:ads_branch_condition}
\end{equation}
rather than by assigning a universal sign to the root in
Eq.~\eqref{eq:f(r)}. The thermodynamic properties of interest are encoded
in the behavior of the metric function at the event horizon, and therefore
our analysis will focus on quantities evaluated at $r=r_+$. The effective coupling $\hat{\alpha}_n$ encodes the strength of the pure Lovelock term of order $n$ and is defined by
\begin{align}
\hat{\alpha}_n =
\begin{cases}
-1/\ell^2, & n = 0, \\[10pt]
1, & n = 1, \\[10pt]
\alpha_n \displaystyle\prod_{i=3}^{2n} (d - i), & n \ge 2.
\end{cases}
\end{align}
Here, $\alpha_n$ denotes the coupling constant multiplying the $n$th-order
Lovelock density in Eq.~\eqref{eq:action}. For $n=1$ we have adopted the
standard Einstein normalization. For $n\geq2$,
the rescaled coupling $\hat{\alpha}_n$ absorbs the dimension-dependent
combinatorial factors arising from the Lovelock tensor in spherical
symmetry. In the vacuum limit $M=Q=0$, Eq.~\eqref{eq:integrated_lovelock} gives
\begin{equation}
\hat{\alpha}_n
\left(-\frac{1}{L_{\rm eff}^2}\right)^n
=
-\frac{1}{\ell^2}.
\end{equation}
Consequently, the effective AdS radius satisfies
\begin{equation}
L_{\rm eff}^{2n}
=
(-1)^{n+1}\hat{\alpha}_n\,\ell^2.
\label{eq:Leff}
\end{equation}
Thus $\ell$, which is introduced through the cosmological constant,
need not coincide with the physical asymptotic AdS curvature radius for
$n>1$. A real AdS vacuum requires
$(-1)^{n+1}\hat{\alpha}_n>0$, together with the choice of algebraic
branch satisfying Eq.~\eqref{eq:ads_branch_condition}. Equivalently,
$\hat{\alpha}_n>0$ for odd $n$, whereas $\hat{\alpha}_n<0$ for even
$n$. In Lovelock theories, the $n$th-order term makes a nonzero
contribution to the local field equations only in dimensions
$d\geq 2n+1$; in $d=2n$ it is topological. In the absence of a
cosmological constant, pure Lovelock gravity has Lovelock-flat vacuum
solutions in its odd critical dimension $d=2n+1$, generalizing the
well-known three-dimensional Einstein-gravity result. These vacua are
not necessarily Riemann-flat, except for $n=1$. When a cosmological
constant is introduced, the Lovelock curvature is instead fixed by the
cosmological constant, and BTZ-like black-hole solutions can arise in
higher-order Lovelock gravity \cite{Dadhich:2012cv}.

Before proceeding further, it is worthwhile mentioning the units being
used. Here and throughout the paper, $\ell$ denotes the cosmological
length scale defined through Eq.~\eqref{eq:lambda}, with $[\ell]=L$,
while $L_{\rm eff}$ denotes the effective asymptotic AdS curvature
radius. The metric function $f(r)$ is dimensionless. For pure Lovelock
gravity of order $n\geq2$, the effective coupling has dimensions
$[\hat{\alpha}_n]=L^{2n-2}$. In geometrized natural units
$\hbar=c=G_d=1$, the geometric mass and charge parameters appearing in
the metric scale as $[M]=L^{d-3}$ and $[Q]=L^{d-3}$, so that each term
inside the parentheses in Eq.~\eqref{eq:f(r)} has dimensions
$L^{-2n}$. In particular,
$\left[1/(\hat{\alpha}_n\ell^2)\right]=L^{-2n}$, and hence the quantity
raised to the power $1/n$ has dimensions $L^{-2}$. The overall factor
$r^2$ then ensures that $f(r)$ is dimensionless.

Imposing the horizon condition $f(r_+)=0$ (equivalently $g^{rr}=0$) yields the mass parameter as a function of the horizon radius and the electrostatic potential,
\begin{align}
M &= \frac{\Sigma_{d-2} (d-2) r_+^{d-1}}{16 \pi }
\left[
\frac{\hat{\alpha}_n}{r_+^{2n}} + \frac{1}{\ell^2}
+ \frac{2(d-3) \Phi^2}{(d-2)r_+^2}
\right].
\end{align}
Here, $\Phi$ denotes the electrostatic potential, defined as the difference in the electric potential between spatial infinity and the event horizon, with the gauge chosen such that the potential vanishes at infinity. It is given by
\begin{align}
\Phi = \frac{4\pi Qr_+^{3-d}}{(d-3)\Sigma_{d-2}}.
\end{align}
In the Euclidean (finite-temperature) formulation $\Phi$ is the boundary datum conjugate to electric charge $Q$. Equivalently, $\Phi$ may be interpreted as the properly normalized value of the gauge potential at the horizon relative to infinity, ensuring regularity of the Euclidean solution \cite{PhysRevD.60.064018, PhysRevD.15.2752}. This identification makes explicit that variations of the on-shell action at fixed $\Phi$ yield the thermodynamic conjugate charge $Q$ in the first law. In the grand-canonical ensemble, $\Phi$ is held fixed. In what follows, the grand-canonical ensemble will be used frequently, as it allows one to treat families of black holes as parametrized by $(r_+, \Phi)$ and to track multiple solution branches at a given temperature in a transparent manner. 

The Hawking temperature follows from the regularity of the Euclidean section (or, equivalently, from the surface gravity at the outer horizon),
\begin{align} \label{eq:TempChargedAdS}
T= \frac{f'(r)|_{r=r_+}}{4\pi} &= \frac{d - 2n - 1}{4 \pi n \, r_+}
+ \frac{d - 1}{4 \pi n \hat{\alpha}_n \ell^2} \, r_+^{2n-1}
- \frac{(d-3)^2 \Phi^2}{2 \pi (d-2) n \hat{\alpha}_n} \, r_+^{2n-3}.
\end{align}
The first term in Eq.~\eqref{eq:TempChargedAdS} represents the
Schwarzschild-like contribution, scaling as the inverse of the outer
horizon radius; the second term represents the AdS curvature
contribution, which grows with the outer horizon radius; and the last
term represents the electromagnetic contribution, controlled by
$\Phi$, with a different power of $r_+$ that may increase or decrease
depending on the Lovelock order $n$. The competition among these terms
can render $T(r_+)$ non-monotonic. When $T(r_+)$ develops a positive
minimum $T_{\min}$, temperatures above $T_{\min}$ can correspond to two
solutions for $r_+$, representing the small- and large-black-hole
branches. This two-branch structure is not universal; in particular,
in the odd critical dimension $d=2n+1$, the Schwarzschild-like term
vanishes and the positive-temperature branch may be monotonic. Where
the two branches exist, their local thermal stability at fixed
potential may be diagnosed from the slope of $T(r_+)$, provided that $T>0$ and $dS/dr_+>0$.

Finally, the entropy follows from the first law of black hole thermodynamics. Using $dM=T\,dS+\Phi\,dQ$ and the temperature above, one obtains \cite{PhysRevD.74.064001, Khuzani:2022lqd} 
\begin{align} \label{eq:entropypureLovelock}
S =\int_0^M T ^{-1}dM &= \frac{(d-2)\Sigma_{d-2} \hat{\alpha}_n \, n \, r_+^{d-2n}}{4  (d-2n)}.
\end{align}
The above expression for the entropy can also be obtained by the Wald entropy expression given in Eq.(23) of \cite{Kolekar:2011bb}  and exploiting the fact that the spatial cross-section of the event horizon of a spherically symmetric black hole in \(d\)-dimensional spacetime is a \((d-2)\)-dimensional maximally symmetric manifold, whose Riemann tensor is given by
\begin{align}
R^{ij}{}_{kl}
=
\frac{1}{r_+^2}
\left(\delta^i_k\delta^j_l-\delta^i_l\delta^j_k\right).
\end{align}
As expected in pure Lovelock theories, the entropy differs from the area law and scales as $r_+^{d-2n}$. However, in the critical dimensions \(d = 2n+1\) and \(d = 2n+2\), the entropy scales as \(S \sim r_+\) and \(S \sim r_+^2\), respectively, independent of the Lovelock order \(n\). This remarkable feature, known as the thermodynamic universality of pure Lovelock black holes, implies that the functional dependence of the entropy on the horizon radius remains unchanged across different Lovelock orders in the critical dimensions, thereby exhibiting an Einstein-like universality in the thermodynamic sector despite the underlying higher-curvature dynamics. The denominator $(d-2n)$ indicates that the expression in Eq.\eqref{eq:entropypureLovelock} is valid in the dynamical regime $d>2n$, where the Lovelock term of order $n$ contributes nontrivially to the equations of motion and the associated black-hole thermodynamic quantities are defined through the resulting dynamical solutions; at $d=2n$, the Lovelock density reduces to a topological Euler term and no longer contributes to the local gravitational dynamics. Together, $T(r_+)$ and $S(r_+)$ provide the basic input for the analysis of stability and phase structure in the charged AdS pure Lovelock black hole family.

Having discussed the required thermodynamic quantities of static spherically symmetric pure Lovelock black holes, we now discuss the Hawking--Page transition, the temperature at which it happens, and how this temperature is related to the minimum temperature. The corresponding result for the case of general relativity was first observed in \cite{PhysRevD.102.104011}, wherein it was referred to as a novel dual relation. 

\section{Hawking--Page transition and the novel duality} \label{sec:temp_relation}

The interplay between different contributions to the temperature in Eq.\eqref{eq:TempChargedAdS}, arising from the Schwarzschild-like, AdS curvature, and (when present) electromagnetic sectors, generically induces a non-trivial structure in $T(r_+)$, allowing for the emergence of a minimum temperature. The minimum temperature encodes a geometric obstruction to the existence of equilibrium black hole solutions in a fixed temperature ensemble. Using the temperature shown in Eq.\eqref{eq:TempChargedAdS} for the electromagnetically chargeless case, $\Phi=0$, and the relation between pressure and AdS length given by
\begin{align} \label{eq:pressureAdSlength}
    P=\frac{(d-1)(d-2)}{16\pi \ell^2},
\end{align}
the minimum temperature is obtained as
\begin{align} \label{eq:minTemp}
    T_0 = \frac{d-2n-1}{2\pi(2n-1)} \bigg( \frac{(2n-1)16\pi P}{(d-2n-1)(d-2)\hat{\alpha_n}}\bigg)^{1/2n}.
\end{align}
The minimum temperature $T_0$ signals a saddle-node bifurcation in the space of solutions. For temperatures $T > T_0$, the equations of motion admit two distinct black hole branches: a small black hole with $r_h \ll l$, which has negative specific heat and is therefore locally thermodynamically unstable, and a large black hole with $r_h \gtrsim l$, which has positive specific heat and is locally stable. At $T = T_0$, these two branches merge, and for $T < T_0$, no real, positive horizon radius exists, so the spacetime cannot support an equilibrium black hole configuration in the fixed temperature ensemble, and the only solution is thermal AdS. The mere existence of black hole solutions does not determine the dominant phase; instead, the Gibbs free energy governs which phase dominates, as we discuss in more detail below. In particular, at a higher temperature $T_{\text{HP}} > T_0$, the Hawking--Page phase transition occurs, where the free energy of the large black hole equals that of thermal radiation in AdS. For $T_0 < T < T_{\text{HP}}$, thermal AdS is globally preferred, while the large black hole is only metastable and the small black hole is both subdominant and unstable. For $T > T_{\text{HP}}$, the large black hole becomes the globally dominant phase, whereas thermal AdS becomes metastable; the small black hole branch continues to exist but remains thermodynamically unstable.

The structure of Eq.\eqref{eq:minTemp} reveals that $T_0$ is governed by a delicate balance between dimensionality $d$ and the pure Lovelock theory parameter $n$, with the combination $d-2n-1$ playing a critical role. This factor effectively measures how far the theory is from a degeneracy point where the thermodynamic description itself breaks down. As $d \to 2n+1$, the minimum temperature is driven toward zero or becomes ill-defined, indicating that the notion of a thermally stable black hole ceases to exist in that limit. Thus, $T_0$ is a diagnostic of the underlying gravitational theory, and it encodes how higher-curvature interactions constrain the very phase space of admissible black hole geometries.

The thermodynamic behavior of asymptotically Anti-de Sitter (AdS) black holes reveals the rich phase structure, most notably the Hawking--Page (HP) phase transition, the phase transition between thermal radiation in AdS and the stable large black hole configuration \cite{Hawking:1982dh}. The  HP phase transition can be interpreted as a confinement/deconfinement transition in the context of gauge/gravity duality, thereby enhancing its physical significance. The sign and relative magnitude of the Gibbs free energy determine the globally preferred phase: thermal AdS or the black hole. Therefore, computing the Gibbs free energy of the system is a crucial step in identifying the occurrence of the HP phase transition and characterizing the associated thermodynamic stability. For the electromagnetically neutral case, the Gibbs free energy
reduces to
\begin{align}
G &= M - TS  \\
&= \frac{\Sigma_{d-2} (d-2)r_+^{d-1}}{16\pi }
\left[
\frac{\hat{\alpha}_n}{r_+^{2n}} 
+ \frac{16\pi P}{(d-1)(d-2)}
-\bigg(
\frac{d-2n-1}{ n r_+}
+ \frac{16\pi P}{ n \hat{\alpha}_n (d-2)}r_+^{2n-1} \bigg) \frac{\hat{\alpha}_n n r_+^{1-2n}}{d-2n} \right].
\end{align}

The HP phase transition is characterized by the vanishing of the Gibbs free energy, $G=0$, which marks the phase transition between thermal AdS space and a stable black hole configuration. Solving the condition $G=0$ for the horizon radius $r_+$ and substituting back into the temperature expression yields the Hawking--Page temperature. It depends explicitly on the spacetime dimension $d$, the theory parameter $n$, and the thermodynamic pressure $P$, and is obtained as
\begin{align} \label{eq:HawkPageTemp}
    T_{\text{HP}} = \frac{(d-2n)}{2\pi(2n-1)}\bigg( \frac{16\pi(2n-1)P}{(d-1)(d-2)\hat{\alpha}_n} \bigg) ^{1/2n}.
\end{align} 
The above expression, Eq.\eqref{eq:HawkPageTemp}, encodes how the onset of black hole dominance over thermal radiation is controlled by both geometric and higher-curvature contributions in AdS spacetime. We postpone the discussion of other features of the Gibbs free energy to the next section.

A remarkable dual relation emerges when comparing the HP phase transition temperature with the minimum temperature $T_0$ of the AdS black hole branch. By relating Eq.~\eqref{eq:HawkPageTemp} with the corresponding expression for $T_0$, one finds that the minimum temperature in $(d+1)$ dimensions is directly tied to the HP phase transition temperature in $d$ dimensions through
\begin{align} \label{eq:novelDualrelationneutral}
    T_0(d+1) = T_{\text{HP}}(d)\,\bigg(\tfrac{d-2}{d-2n}\bigg)^{1/2n}\, .
\end{align}
The above Eq.\eqref{eq:novelDualrelationneutral} suggests that in general relativity ($n=1$), the minimum temperature in $d+1$ dimensions is the same as the HP phase transition temperature in $d$ dimensions. In pure Lovelock gravity, the dual relation gets a correction depending on the dimension of the spacetime and the order of the Lovelock Lagrangian, but is still independent of all thermodynamic parameters. What is striking about this relation is that it does not merely compare two temperatures within the same thermodynamic ensemble, but instead ties together two different physical thresholds across different spacetime dimensions. The HP phase transition temperature $T_{\mathrm{HP}}$ encodes a \emph{global} phase transition (thermal AdS $\leftrightarrow$ AdS black hole) determined by the crossing of free energies, whereas $T_0$ is the \emph{minimum} black-hole temperature (the cusp of $T(r_h)$) so that for $T<T_0$ there is no \emph{equilibrium} AdS black-hole solution in the fixed temperature ensemble (and for $T>T_0$ one typically finds a small, locally unstable branch and a large, locally stable branch). The fact that $T_0(d+1)$ is algebraically determined by $T_{\text{HP}}(d)$ suggests that the onset of black hole \emph{existence} in a given dimension is dual to the onset of black hole \emph{dominance} in a lower dimension. This implies that the global phase structure in $d$ dimensions already ``knows'' about the stability edge of the theory in $(d+1)$ dimensions.

The HP phase transition temperature $T_{\rm HP}$ is fixed by an \emph{off-shell} comparison of Euclidean actions (or Gibbs free energies) between two phases, while $T_0$ is fixed by an \emph{on-shell} turning point condition along the black hole family, $({\partial T}/{\partial r_+})_{r_0}=0$. The duality, therefore, relates an action-level global criterion in $d$ dimensions to a local geometric criterion in $(d+1)$ dimensions, indicating that the information contained in the free-energy landscape of the lower-dimensional theory is reorganized into the branch-structure geometry of the higher-dimensional theory. In particular, since $T_0$ is the temperature at which the radiation and small/large black hole branches meet (and the specific heat changes sign), the map suggests that the HP phase transition point in $d$ dimensions is the dimensional precursor of the branch point at which the canonical black hole branch nucleates in $(d+1)$ dimensions.

Although both $T_{\rm HP}$ and $T_0$ separately depend on the AdS scale (or pressure) through the same overall scaling, that dependence cancels out of the ratio implied by the above relation, leaving behind a purely dimension-theoretic deformation controlled by $n$. The scaling factor $\big((d-2)/(d-2n)\big)^{1/2n}$ isolates the role of the pure Lovelock order $n$ in a way that is insensitive to the pressure and temperature, indicating that this duality is geometrical. The independence from pressure, temperature, horizon radius, etc., indicates that the duality is not driven by the details of the equation of state but rather by the symmetry of the pure Lovelock theory Lagrangian and how the order $n$ reorganizes the radial dependence of $T(r_+)$ across dimensions.

The correction factor also encodes a consistency domain that is physically meaningful. The condition $d-2n>0$ is precisely the condition under which the pure Lovelock term contributes dynamically, and an AdS black hole branch with a genuine minimum temperature exists. Thus, the dual relation automatically ``knows'' about the admissible dimensional window of the theory: when $d\to 2n$, the factor diverges, signaling the breakdown of the standard thermodynamic branch structure (and, correspondingly, of the interpretation of $T_0$ as an existence threshold). In the Einstein limit $n=1$, the factor reduces to unity, and the duality becomes exact, $T_0(d+1)=T_{\rm HP}(d)$, whereas for $n>1$ it measures the precise displacement between the dimensional images of global and local thresholds induced by higher-curvature interactions.

From a holographic perspective, the cross-dimensional relation looks suggestive because $T_{\rm HP}(d)$ is the temperature of confinement/deconfinement for the putative $d$-dimensional boundary theory, while $T_0(d+1)$ is a bulk obstruction temperature below which no black hole saddle exists in the $(d+1)$-dimensional gravity theory. The map, therefore, links a lower-dimensional boundary transition scale to a higher-dimensional bulk existence bound, hinting at a recursive structure in which the data controlling saddle dominance in one dimension reappears as the data controlling saddle availability one dimension higher. This interpretation is reinforced by the fact that the mapping does not involve any additional thermodynamic variables, but only $d$ and $n$, making it a candidate universality statement about AdS black holes in pure Lovelock classes rather than an artifact.

In this sense, the shift $d\to d+1$ paired with a rescaling of temperature resembles a renormalization-like flow in dimension, but with an important refinement: the ``flow'' exchanges the criterion $G=0$ (dominance) with the criterion $\partial_{r_+}T=0$ (branch endpoint). This relation further suggests that phase transitions and extremality/turning-point bounds are different projections of a single, dimensionally covariant principle governing AdS black hole thermodynamics, with the pure Lovelock order $n$ controlling how sharply these projections are separated as one moves across dimensions.

\begin{figure}[H]
    \centering
    \includegraphics[width=.92\textwidth]{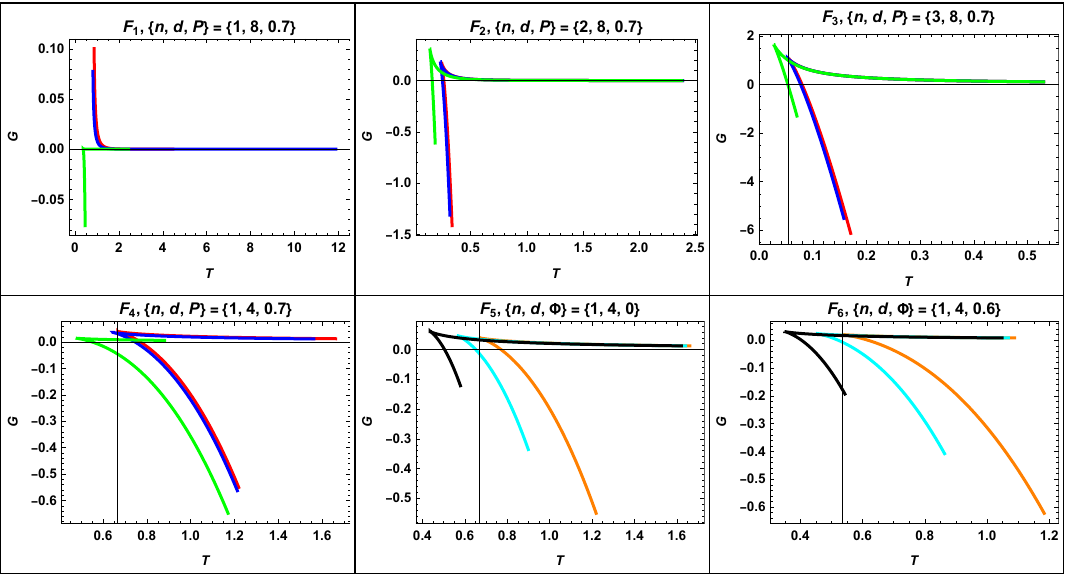}
     \captionsetup{margin=1cm, font=small}
    \caption{The above figure shows the plots for the Gibbs free energy vs temperature. In panels $F_1, F_2, F_3,$ and $F_4$, the colours $\{\text{Red, Blue, Green}\}$ represent the potential $\Phi$ with values $\{0, 0.25, 0.7\}$. In figures $F_5$ and $F_6$, the colours $\{\text{Orange, Cyan, Black}\}$ represent pressure values of $\{0.7, 0.5, 0.3\}$.} 
    \label{GibbsFreeEnergy:1}
\end{figure}

\section{HP transition for pure Lovelock black holes with electric charge} \label{sec:curvature}

In the Euclidean formulation of black-hole thermodynamics, the gravitational partition function is evaluated semiclassically by a
saddle-point approximation, $Z(T,\Phi)\simeq
\exp\!\left[-I_E^{\rm ren}\right]$, where $I_E^{\rm ren}$ is the
renormalized Euclidean on-shell action of the relevant saddle. The
Euclidean time coordinate is periodic with period
$\beta=1/T$ \cite{GibbonsHawking1977}. Imposing Dirichlet boundary
conditions on the gauge potential, so that the electric potential
$\Phi$ is held fixed at the boundary, defines the grand-canonical
ensemble. Here $\Phi$ is understood as the gauge-invariant potential
difference between the asymptotic boundary and the horizon; in a
regular Euclidean gauge one conventionally sets $A_\tau(r_+)=0$. In
this ensemble the corresponding thermodynamic potential is the grand
potential, commonly referred to as the Gibbs free energy in the
black-hole literature. For fixed gauge potential, no additional
electromagnetic Legendre boundary term is required. By contrast, such
a boundary term must be included when passing to the canonical
ensemble in which the electric charge is held fixed
\cite{BradenEtAl1990,HawkingRoss1995}.

The semiclassical relation between the Euclidean on-shell action and
the thermodynamic potential is a general feature of gravitational
thermodynamics. What is specific to Lovelock gravity is the
gravitational boundary term required for a well-posed Dirichlet
variational principle. For the pure Lovelock theory considered here,
the ordinary Gibbons--Hawking--York term must be replaced by the
corresponding generalized Gibbons--Hawking--Myers boundary term
associated with $\mathcal{L}_n$ \cite{Myers1987,MiskovicOlea2007}. In
asymptotically AdS spacetimes, the resulting on-shell action must
furthermore be appropriately renormalized. In the Dirichlet
formulation, this may be accomplished by intrinsic boundary
counterterms \cite{MiskovicOlea2007,Yale2011}; alternatively, one may
employ the Kounterterm regularization prescription
\cite{KofinasOlea2007}.

Once the variational principle and the asymptotic regularization are
properly implemented, the renormalized Euclidean on-shell action yields
the grand-canonical thermodynamic potential through the quantum
statistical relation
\cite{MyersSimon1988,MiskovicOlea2011,AranguizKuangMiskovic2016}
\begin{equation}
G
=
\frac{I_E^{\rm ren}}{\beta}
=
M-TS-\Phi Q,
\label{eq:GibbsChargedAdS}
\end{equation}
where $M$ is the renormalized conserved mass defined in the same
renormalization scheme as $I_E^{\rm ren}$, $S$ is the Wald entropy,
which in Lovelock gravity reduces to the Jacobson--Myers entropy
\cite{JacobsonMyers1993}, and $T$ and $\Phi$ are held fixed. This
agrees with the thermodynamic potential obtained from the first law,
provided that the mass and Euclidean action are defined in the same
renormalization scheme. When the thermodynamic potential is
reconstructed by integrating the first law, an additive vacuum-energy
normalization may remain. A Hawking--Page transition, when present, may therefore be identified
by comparing the renormalized Euclidean actions of the black-hole and
thermal-AdS saddles subject to the same boundary data
\cite{HawkingPage1983}. Here thermal AdS is understood to belong to the
same asymptotic Lovelock-AdS vacuum branch as the black-hole solution
\cite{CamanhoEdelstein2013}. The coexistence condition is $\Delta G
\equiv
G_{\rm BH}-G_{\rm AdS}
=
0.$
If the Gibbs free energy of thermal AdS is chosen as the reference
value, $G_{\rm AdS}=0$, this condition reduces to
$G_{\rm BH}=G_{\rm AdS}=0$.

In the ensemble with fixed $\Phi$ the phase transition of the charged black hole, described by line element Eq.\eqref{eq:purLovelockMetric}, the Gibbs free energy, Eq.\eqref{eq:GibbsChargedAdS}, becomes
\begin{align}
G  &= \frac{\Sigma_{d-2} (d-2)r_+^{d-1}}{16\pi }
\left[
\frac{\hat{\alpha}_n}{r_+^{2n}} 
+ \frac{16\pi P}{(d-1)(d-2)}
+ \frac{2(d-3) \Phi^2}{(d-2)r_+^2}
\right]  \nonumber \\
&\quad 
-\left[
\frac{d-2n-1}{4\pi n r_+}
+ \frac{4P}{ n \hat{\alpha}_n (d-2)}r_+^{2n-1}
- \frac{(d-3)^2  \Phi^2}{2\pi (d-2)n \hat{\alpha}_n}r_+^{2n-3}
\right] \nonumber \\
&\quad
\times \frac{(d-2)\Sigma_{d-2}\hat{\alpha}_n n r_+^{d-2n}}{4(d-2n)}
- \frac{\Phi^2 (d-3)\Sigma_{d-2}}{4\pi} r_+^{d-3} \label{eq:GibbsFreeExplicitChargAdS}.
\end{align}

The Gibbs free energy at fixed temperature and electrostatic potential provides a direct diagnostic of global thermodynamic dominance in the grand canonical ensemble. For illustration purposes, we show the plot for the Gibbs free energy in Fig.\ref{GibbsFreeEnergy:1}. In each panel, lower branches describe physically preferable and stable large black holes in equilibrium, while $G=0$ determines the Hawking--Page (HP) temperature $T_{\rm HP}$ at which the dominant saddle exchanges between thermal AdS (taken to have vanishing Gibbs free energy) and the AdS black hole. The plots exhibit the standard picture of an HP-type transition: at sufficiently low temperature, the relevant black-hole branch carries nonnegative free energy and is therefore subdominant, whereas increasing $T$ drives $G$ downward and the black-hole free energy becomes negative, signaling that the black hole becomes the globally preferred configuration. Importantly, the multi-branch structure visible in all panels indicates the presence of competing black-hole saddles (typically interpreted as small/large black-hole branches in AdS), and the cusp/turning-point behavior corresponds to the boundary of local thermodynamic stability, where the heat capacity changes sign. Consequently, the HP transition identified by $G=0$ is to be distinguished from intra-black-hole first-order transitions (swallowtail-like behavior); here, the principal effect is the exchange of dominance with the thermal AdS background, shaped by the existence and stability of multiple black-hole branches.

Varying the theory parameter $n$ produces a pronounced deformation of the Gibbs free energy and a systematic displacement of HP phase transition temperature, demonstrating that the confinement/deconfinement-like transition scale is strongly theory dependent. As $n$ increases, the free-energy curves generically shift leftward, and the descent of $G(T)$ with temperature becomes steeper, so that the crossing $G=0$ occurs at a lower temperature; equivalently, larger $n$ enhances the thermodynamic preference for the black-hole saddle once thermal excitations are turned on. One can interpret this behavior to be originated from the modified thermodynamic scaling introduced by higher-order Lovelock curvature terms. As the Lovelock order $n$ increases, the relative balance between the mass term and the thermal contribution $TS$ in the Gibbs free energy changes in such a way that the free energy decreases more rapidly with temperature. Consequently, the condition $G=0$ is reached at a lower temperature, indicating that higher-curvature effects enhance the thermodynamic preference for the black-hole phase over the thermal background. Physically, this behavior reflects the stronger role of higher-curvature interactions in pure Lovelock gravity. As the Lovelock order $n$ increases, the gravitational dynamics near the horizon become increasingly governed by higher-order curvature terms, which effectively lower the thermodynamic cost of sustaining a black-hole geometry in AdS spacetime. Consequently, the black-hole phase becomes competitive with the thermal AdS background at lower temperatures, leading to an earlier HP phase transition and a stronger thermodynamic preference for the black-hole saddle once thermal excitations are introduced. Turning on the electrostatic potential $\Phi$ further biases the transition by lowering the temperature at which the transition occurs, and changing the region in which charged AdS black holes are globally stable/unstable. In particular, the plots shift downward if we increase the electrostatic potential. However, we see that if we keep the pressure fixed and keep on increasing $\Phi$, then after a certain value of $\Phi$, the HP transition becomes unlikely to happen as we discussed above. 

For general relativity, $n=1$, the Gibbs free energy Eq.\eqref{eq:GibbsChargedAdS} vanishes at the Hawking temperature given by 
\begin{equation} \label{eq:HawkTempGR}
    T_{\text{HP}}^{\text{GR}} = \sqrt{\frac{4P(d-2-2(d-3)\Phi^2)}{\pi (d-1)}}.
\end{equation}
Here, we assume that $d - 2 > 2(d - 3)\Phi^2$ to ensure the reality of $T_{\text{HP}}^{\text{GR}}$. In the case of a sufficiently large electric potential such that $d - 2 < 2(d - 3)\Phi^2$,  we do not get any real root of the equation given by the vanishing of the Gibbs free energy. Therefore, the HP phase transition does not occur in Einstein theory at a large electrostatic potential in the grand canonical ensemble.
We note from Eq.\eqref{eq:TempChargedAdS} that the above HP phase transition temperature Eq.\eqref{eq:HawkTempGR} is related to the minimum temperature of the AdS spacetime in dimensions $d$ as
\begin{align} \label{eq:novelGRCharged}
    T_{\text{0}}^{\text{GR}}(d) =  T_{\text{HP}}^{\text{GR}} (d) \sqrt{\frac{(d-1)(d-3)}{(d-2)^2}}.
\end{align}
The above relation, Eq.~\eqref{eq:novelGRCharged}, remains valid independently of the electric potential and pressure, provided that both quantities are well defined. In the limit of large dimensions the square root factor in Eq.\eqref{eq:novelGRCharged} tends to 1. Therefore, the minimum temperature and the HP phase transition temperature are close to each other, and we have a very small region of metastable black hole branch. The HP phase transition temperature in $(d+1 )$ dimensions is related to the minimum temperature of $d$ dimension as
\begin{align} \label{eq:noveldualGRCharged}
     T_{\text{HP}}^{\text{GR}} (d,\Phi)= T_{\text{0}}^{\text{GR}} \bigg(d+1,\frac{\sqrt{(d-1)(d-3)}}{d-2}\Phi\bigg) .
\end{align}
The above generalized duality relation, Eq.~\eqref{eq:noveldualGRCharged}, establishes a mapping between thermodynamic quantities in adjacent spacetime dimensions within the grand canonical ensemble. In particular, it shows that the HP phase transition temperature in $d$ dimensions at fixed electric potential $\Phi$ is equal to the minimum temperature in $(d+1)$ dimensions, provided the potential is rescaled by an appropriate dimension-dependent factor. This demonstrates that a change in spacetime dimensionality can be compensated by a correlated transformation of the gauge potential, leaving the thermodynamic threshold invariant. Physically, this relation reflects an equivalence between geometric degrees of freedom (spacetime dimension) and gauge degrees of freedom (electric potential), indicating that black hole stability is governed by a combined parameter space rather than independent variables. The duality, therefore, organizes black hole states into equivalence classes characterized by invariant transition temperatures under this joint transformation. In the large-dimension limit ($d \to \infty$), the rescaling factor approaches unity, implying that the distinction between the two temperatures diminishes. In this regime, the duality becomes asymptotically trivial, and the thermodynamic behavior in successive dimensions converges, leading to a simplified phase structure.

\section{HP transition from thermodynamic geometry} \label{sec:curvature}

In the framework of thermodynamic geometry, one equips the space of equilibrium states with a Riemannian metric whose structure is determined by the underlying thermodynamic potential. The thermodynamic geometric formulation provides a powerful tool to probe phase structure and critical phenomena, as geometric invariants constructed from the metric can capture information about the underlying interactions and correlations in the system. In the grand canonical ensemble with fixed electrostatic potential $\Phi$, the appropriate thermodynamic potential is the Legendre transform of the Helmholtz free energy,
\begin{equation}
\mathcal{G}(T,V,\Phi)=F(T,V,Q)-\Phi Q,
\end{equation}
whose natural variables are $\{T,V,\Phi\}$ and which satisfies
\begin{equation}
d\mathcal{G}=-S\,dT - P\,dV - Q\,d\Phi.
\end{equation}

In this formulation, the thermodynamic metric is constructed from the Hessian of $\mathcal{G}$ with respect to the variables $\{T, V, \Phi\}$. In the following, we focus on the corresponding two-dimensional line element on the constant-$\Phi$ hypersurface. Taking temperature and volume as fluctuation coordinates, and $\mathcal{G}$ as the thermodynamic potential, the thermodynamic line element on the hypersurface of constant $\Phi$,
\begin{align} \label{eq:generalthermometricdef}
    \Delta l^2 = \frac{1}{T} \Delta T \Delta S -  \frac{1}{T} \Delta P \Delta V
\end{align}
can be written as\footnote{Here we substitute 
\begin{align*}
    \Delta S &= (\partial_TS)\Delta T + (\partial_V S)\Delta V \\
    \Delta P &= (\partial_T P)\Delta T + (\partial_V P) \Delta V
\end{align*}
with $S=-(\partial_T\mathcal G)_{V,\Phi}$ and $P=(\partial_V\mathcal{G})_{T,\Phi}$ in Eq.\eqref{eq:generalthermometricdef}.} \cite{PhysRevD.100.124033}
\begin{align} \label{eq:thermodynamicLineElement}
dl^2=\frac{C_V}{T^2}\,dT^2-\frac{1}{T}\left(\frac{\partial P}{\partial V}\right)_{T,\Phi}dV^2\, .
\end{align}
where $C_V = T\left(\partial_T S\right)_V$ is the heat capacity at constant thermodynamic volume $V$. The coefficient of $dT^2$ is governed by the heat capacity, reflecting thermal fluctuations, while the coefficient of $dV^2$ involves the isothermal derivative of pressure at fixed electrostatic potential, encoding the mechanical response of the system under volume fluctuations. Consequently, the metric components are directly related to thermodynamic response functions. The positivity of $C_V$ and $-\left(\partial P/\partial V\right)_{T,\Phi}$ for the large black hole branch ensures that the metric components are positive, corresponding to positive thermodynamic susceptibilities and hence stability against local fluctuations.

We now turn to the thermodynamic geometry and analyze the normalized Ruppeiner scalar \( R_N \), defined as the heat capacity at constant volume multiplied by the scalar curvature of the thermodynamic metric \cite{PhysRevD.100.124033}, for charged, large AdS black holes in the extended phase space. The normalized scalar curvature for the above thermodynamic metric in Eq.\eqref{eq:thermodynamicLineElement}, assuming $C_V$ to be constant, is obtained as
\begin{align}  \label{eq:chargedRuppiener}
R_N=\frac{1}{2}
- \frac{1}{2
\left(
1 + \frac{(d - 2n - 1)\, r_+^{-1}}{2\pi (1 - 2n)\, T}
- \frac{(d - 3)^2\,\, \Phi^2\, r_+^{2n - 3}}{n \pi \hat{\alpha}_n (d - 2)(1 - 2n)\, T}
\right)^{2}
}.
\end{align}
For the chargeless black hole case in the pure Lovelock theory, the above relation, Eq.\eqref{eq:chargedRuppiener}, at the Hawking-Page temperature shown in Eq.\eqref{eq:HawkPageTemp} becomes 
\begin{align} \label{eq:chargeLessRuppieneraab}
R_N= -\frac{(d-2n-1)(d-2n+1)}{2}
\end{align}
The above expression of the scalar curvature, Eq.\eqref{eq:chargeLessRuppieneraab}, depends only on the spacetime dimension and the order of the theory considered. Therefore, the normalized Ruppiener scalar is a constant of the HP phase transition in pure Lovelock theories. The constancy of the normalized Ruppeiner scalar at the HP phase transition indicates that, despite the discontinuous change in the dominant thermodynamic phase, from thermal AdS to a large black hole, the transition occurs at a universal, dimension-dependent value of the normalized microscopic interaction strength. Since $R_N$ encodes both the nature and the effective strength of microscopic interactions, its constancy across the transition also suggests that the underlying microstructure reorganizes with an approximately universal correlation scale or effective correlation volume. In particular, since $d \geq 2n+1$, the Ruppeiner scalar curvature at the HP phase transition given in Eq.\eqref{eq:chargeLessRuppieneraab} is negative. The fixed negative sign of $R_N$, determined solely by $(d,n)$, implies that the interactions in the corresponding ensemble are dominantly attractive throughout the transition, rather than being driven by a change in interaction type.

The HP phase transition is understood as a global change in dominance between competing saddles of the gravitational path integral, rather than a local instability in the fluctuation spectrum. The fact that $R_N$ depends only on the discrete parameters $(d,n)$ further suggests a form of universality governed by spacetime dimensionality and higher-curvature theory. This reinforces the view that the transition is controlled by global geometric features of the underlying spacetime rather than by dynamical critical behavior. Consequently, the constancy of $R_N$ provides evidence that the dual field theory undergoes a confinement/deconfinement transition with a universal normalized curvature scalar at the phase transition and is consistent with a first-order phase transition in the large-$N$ limit. The constancy of the normalized Ruppeiner scalar $R_N$ at the HP phase transition further supports the interpretation that the transition is first order rather than critical.

At the HP phase transition, Eq.~\eqref{eq:HawkTempGR}, the normalized Ruppeiner scalar at the HP phase transition in general relativity ($n = 1$) can be written from Eq.~\eqref{eq:chargedRuppiener} as: 
\begin{align} \label{eq:RuppiennerScalarGR}
R_N &= \frac{1}{2} - \frac{(d-2)^2}{2}
= -\frac{(d-3)(d-1)}{2}.
\end{align}
which is dependent only on dimensions of the spacetime (see the Appendix [\ref{AppendicRuppienerGRexplicit}] for a detailed calculation). It is interesting to note that the above expression in Eq.\eqref{eq:RuppiennerScalarGR} is the same as the chargeless case of Eq.\eqref{eq:chargeLessRuppieneraab} with $n=1$. Therefore, in general relativity, provided $d - 2 > 2(d - 3)\Phi^2$,  electrically charged static black holes in AdS spacetime, too, have a constant Ruppeiner scalar at the Hawking-Page transition in a fixed electric potential ensemble. However, for a large electrostatic potential, $\Phi \to \infty$, or when both $\Phi \to \infty$ and $P \to \infty$ simultaneously, no HP phase transition occurs in general relativity(see Eq.\eqref{eq:HawkTempGR} above and Appendix [\ref{Append:bothLargeLimit}]).

We show the plots for the scalar curvature at the HP phase transition, up to $n=3$ in Fig.\ref{RuppienerCurvature:1}, while in Appendix~[\ref{Appen:smallpotentApprox}] and [\ref{Append:bothLargeLimit}] we present an analytical calculation for all pure Lovelock theories. For \( n > 1 \), we find that both the scalar curvature and the relationship between the minimum temperature and the HP phase transition temperature in AdS spacetime generally depend on the pressure and the electrostatic potential. However, in the limit of large pressure and fixed electrostatic potential, i.e., $\Phi^2 P^{-(1-1/n)} \ll 1$, the normalized Ruppeiner scalar asymptotically becomes a constant, the same as the normalized Ruppeiner scalar shown in Eq.\eqref{eq:chargeLessRuppieneraab}, which is independent of pressure, temperature or the horizon radius (see Appendix[\ref{Appen:smallpotentApprox}]). Moreover, in the limit where both the electrostatic potential and pressure tend to infinity, we get the normalized scalar curvature to be dependent only on the dimension of the spacetime and the theory considered for $n\geq 2$. It is given by the following: 
\begin{align} \label{eq:bothPPhilargelimit}
    R_N(P\rightarrow \infty,\Phi\rightarrow \infty) \approx  \frac{1}{2} - \frac{(2n-1)^2(8\pi C^2-(d-3)^2)^2}{2((2n-1)(8\pi C^2-(d-3)^2)+2(d-3)^2)^2},
\end{align}
where $C$ is a constant that depends on only the spacetime dimension and the theory considered (see Appendix[\ref{Append:bothLargeLimit}] ). For the case of general relativity in the same simultanuous large pressure and electrostatic potential limit, the Hawking–Page phase transition remains improbable (see Appendix~\ref{Append:bothLargeLimit}).

\begin{figure}[H]
    \centering
    \includegraphics[width=.92\textwidth]{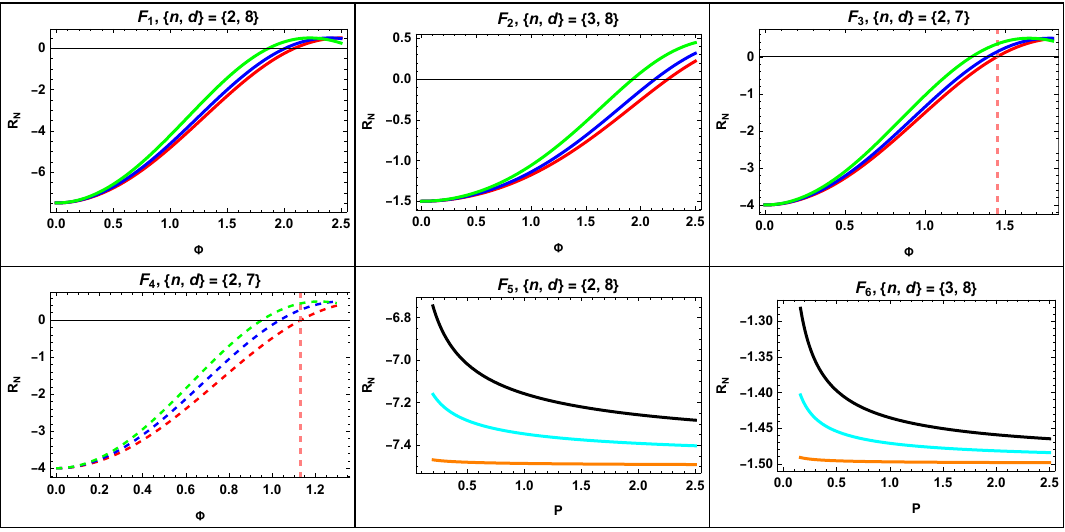}
     \captionsetup{margin=1cm, font=small}
    \caption{The above plots depict the dependence of the normalized Ruppeiner scalar curvature on the electric potential $\Phi$ and pressure $P$. In panels $F_1,F_2,F_3,F_4$ the pressure is kept constant. Solid Colour plots with color $\{\text{Red,Blue, Green}\}$ represent pressure $\{\text{27,23, 17}\}$ while the dashed plots with color $\{\text{Red, Blue, Green}\}$ have pressure $\{\text{10,7,5}\}$. In panels $F_5$, $F_6$ the electric potential is fixed with colors $\{\text{Orange,Cyan, Black}\}$ representing electric potential $\{\text{0.03,0.1, .15}\}$. The pink dashed curves in $F_3$ and $F_4$ describe the constant electrostatic potential at which $R_N$ of the red curves changes sign from negative to positive. } 
    \label{RuppienerCurvature:1}
\end{figure}

In the large electric potential limit, for a fixed pressure, in Eq. \eqref{eq:GibbsFreeExplicitChargAdS}, the $\Phi^2$-enhanced terms given by
\begin{equation} 
\frac{\Sigma_{d-2}(d-3)\Phi^2}{8\pi}\,r_+^{d-3}
+\frac{\Sigma_{d-2}(d-3)^2\Phi^2}{8\pi(d-2n)}\,r_+^{d-3}
-\frac{2\Sigma_{d-2}(d-3)\Phi^2}{8\pi}\,r_+^{d-3}
=
\frac{\Sigma_{d-2}(d-3)\Phi^2}{8\pi}\left(\frac{2n-3}{d-2n}\right)r_+^{d-3}
\label{eq:phi_only_reduction}
\end{equation}
will dominate. Since for the pure Lovelock theory \( d > 2n \), and we work with \( d > 3 \), the prefactor is nonzero; therefore, for vanishing Gibbs free energy in Eq.\eqref{eq:GibbsFreeExplicitChargAdS} one needs
\begin{equation}
r_+^{d-3}\rightarrow 0 \qquad \Longrightarrow \qquad r_+\rightarrow 0.
\end{equation}
Hence, in the strict limit $\Phi\to\infty$ with $P$ fixed, the size of the black hole should go to 0 in order to make the Gibbs free energy vanish. Large black holes typically have $r_+ \sim l$, it will be required for the AdS length also to go to 0. However, $l$ is related to pressure as shown in Eq.\eqref{eq:pressureAdSlength} that we have kept fixed. This suggests that, similar to the case of the Einstein theory, the Hawking--Page transition with large black holes is improbable for a large electrostatic potential even in higher pure Lovelock theories, if the remaining terms are kept fixed. One can also understand this from the expression for the temperature given in Eq.~\eqref{eq:TempChargedAdS}: if one increases only the electrostatic potential, then beyond a certain large value of the electrostatic potential, the black hole can cross the extremal limit, beyond which no horizon exists. Therefore, we restrict our analysis to sufficiently small values of $\Phi$ for which the temperature remains positive.

The equation governing the HP phase transition, i.e., the vanishing of the Gibbs free energy, can be solved analytically up to $n=4$. For illustrative purposes, we present the plots of the analytical expressions for the normalized scalar curvature in Fig.~\ref{RuppienerCurvature:1}. The plots are restricted to a certain range of the electrostatic potential because, for a fixed pressure, values of the electrostatic potential beyond a critical limit render the temperature negative. Consequently, the HP transition is not possible beyond that point. A robust feature common to all panels is that, for the HP phase transition, the Ruppeiner curvature becomes essentially constant after an initial variation, with each curve approaching a nearly horizontal plateau. This suggests that the HP transition selects a universal value of the fluctuation geometry, reminiscent of an effective geometric ``attractor.'' Although, for $n\geq 2$, charge deformations modify $R_N$ at small pressures, the large--$P$ branch along the HP curve is characterized by a universal interaction measure, $R_N$, provided the electrostatic potential is sufficiently small. In this sense, the microstructure of the charged large black hole phase at the HP transition is effectively universal, even beyond the Einstein theory, for the higher order pure Lovelock black holes . This agrees with the discussion around Eq.\eqref{eq:bothPPhilargelimit}. 

In contrast to the case of general relativity, where the Ruppeiner scalar shown in Eq.~\eqref{eq:RuppiennerScalarGR} is always negative, we find that it can also become positive in certain regions of the parameter space for \( n \geq 2 \). A positive Ruppeiner scalar indicates that repulsive interactions dominate the effective microstructure of the system. In such cases, black hole formation can be understood as arising from a balance between repulsive microscopic interactions and the overall gravitational attraction, rather than purely attractive binding. This suggests a more intricate internal structure, where the effective microscopic interactions repulsion is not sufficient to prevent collapse but instead modifies the thermodynamic behavior of the resulting black hole.

We note from panels $F_1$ and $F_2$ (both at $d=8$) that increasing the theory parameter from $n=2$ to $n=3$ decreases the magnitude of the thermodynamic normalized curvature scalar at low electrostatic potential.  For $n=2$ (panel $F_1$), the scalar curvature starts at a large negative value at low $\Phi$, and then the magnitude decreases toward a small, near-zero plateau.  For $n=3$ (panel $F_2$), the low-$\Phi$ scalar curvature value is a bit closer to zero, and increasing $\Phi$, it reaches a positive value with slightly greater slope.  Since negative $R_N$ is typically associated with dominantly attractive effective interactions in the underlying degrees of freedom, the systematic lower negative value of the scalar curvature signifies a weakening of effective attraction when higher-order corrections are turned on.  Importantly, the change is not a mere overall rescaling: the sharp low-$\Phi$ variation visible for $n=2$ becomes milder for $n=3$, suggesting that the higher-order parameter $n$ damps the sensitivity of microscopic correlations to charge-sector deformations at the HP transition.

A complementary trend is revealed by comparing the $n=2$ panels at different dimensions.  In panel $F_1$ ($d=8$), the curvature is negative over a large fraction of the displayed range and approaches a small magnitude plateau from below. By contrast, in panel $F_3$ ($d=7$), the curves cross into the positive-curvature regime at a lower electric potential and attain a maximum positive scalar curvature that exceeds the corresponding value in $F_1$. Thus, lowering the dimension from $d=8$ to $d=7$ (at fixed $n=2$) qualitatively reorganizes the thermodynamic geometry: the HP large-black-hole branch moves from an attractive ($R_N<0$) regime to a repulsive ($R_N>0$) regime at comparably lower electrostatic potential. Furthermore, the less negative value of the scalar curvature in the lower dimension suggests that the decrease in dimensions also makes the interactions less attractive.  This again confirms that the effective micro-interaction encoded by the Ruppeiner geometry is not determined by charge alone, but depends crucially on the dimensionality of the underlying gravitational phase space.  In particular, the near constancy and positivity suggest a repulsion-dominated, weakly $\Phi$-dependent microstructure for the HP large-black-hole phase.

Panel $F_4$ further demonstrates that the saturation behavior persists even when the pressure is slightly decreased to the lower-pressure set (dashed curves). However, if the pressure is sufficiently large, then in the low electrostatic potential regime, we observe a comparatively larger region of negative normalized scalar curvature, indicating a larger attractive correlation region.  While the detailed onset of the plateau is shifted (the crossover occurs at slightly lower $\Phi$ for lower pressures), the qualitative pattern remains unchanged: the curvature approaches an almost constant value and does not exhibit any oscillation or additional structure. This supports the interpretation that the HP phase transition constraint organizes the fluctuation geometry into a simple universal form: a short low-$\Phi$ dependence followed by a pressure-robust positive value.

When the electric potential is fixed, and the cosmological constant is varied through pressure, the Ruppeiner curvature shows a different, yet equally systematic, behavior.  In both panels $F_5$ ($n=2$, $d=8$) and $F_6$ ($n=3$, $d=8$), $R_N$ is negative throughout the plotted pressure range and its magnitude is monotone increasing with $P$, approaching a near-constant value at high $P$.  This implies that along the HP transition, the effective attraction in the microstructure is strongest at high pressure and progressively becomes stronger as the AdS pressure is increased.  The emergence of a high-$P$ plateau again signals a saturation regime in which further tuning of the cosmological constant produces only marginal changes in the fluctuation geometry.

The ordering of curves with respect to $\Phi$ is also consistent across all panels: larger $\Phi$ yields a less negative scalar curvature at fixed $P$. Thus, increasing electric potential drives the system toward weaker effective attraction along the HP curve, while increasing pressure drives it toward stronger effective attraction, and both do so in a smooth, noncritical manner. Moreover, raising $n$ from $2$ to $3$ (comparing $F_5$ with $F_6$) compresses the range of $R_N$ and brings the curves closer together, indicating that higher-order pure Lovelock black holes have a reduced sensitivity of the microstructure to both $P$ and $\Phi$ variations.  In other words, the $n=3$ theory exhibits a more ``rigid'' thermodynamic geometry along the HP phase transition.

Taken together, the plots suggest that the HP transition for charged large black holes selects a remarkably simple thermodynamic-geometric structure: $R_N$ undergoes a short variable response to the control parameter ($\Phi$ in $F_1$--$F_4$, or $P$ in $F_5$--$F_6$), and then settles into a nearly constant plateau.  The plateau value and even the sign of $R_N$ depend on $(n,d)$, revealing that the order of the pure Lovelock theory and dimensionality can qualitatively alter whether the effective micro-interactions are attraction- or repulsion-dominated, but once this regime is reached, the interaction measure becomes largely insensitive to further variations of the intensive variables.  Therefore, along the HP curve, the charged large-black-hole phase exhibits an emergent universality, while the parameters $(n,d)$ control which universality class (attractive vs.\ repulsive) is realized.

\section{Conclusion and Discussion} \label{sec:conclusion}

We investigated the thermodynamic structure of static, spherically symmetric AdS black holes in pure Lovelock theories of gravity, with particular emphasis on a novel dual relationship between characteristic temperatures, as well as on the universality of Ruppeiner scalar curvature at the Hawking–Page (HP) phase transition. The analysis reveals a remarkable interplay among spacetime dimensionality, the presence of charge, and the order of the gravitational theory. In particular, the emergence of universal features across different Lovelock orders suggests the existence of deeper geometric and thermodynamic principles underlying the gravitational system. The main conclusions are summarized below.

For electromagnetically neutral, static, spherically symmetric AdS black holes in pure Lovelock gravity, we have established a duality relation whereby the minimum temperature in $(d+1)$ dimensions differs from the Hawking--Page (HP) transition temperature in $d$ dimensions only by a factor determined by the spacetime dimension and the Lovelock order. In the Einstein limit, this factor reduces to unity, implying an exact equivalence between the minimum temperature in $(d+1)$ dimensions and the HP transition temperature in $d$ dimensions. This result points to the existence of an underlying scaling structure governing the thermodynamic phase behavior of AdS black holes across dimensions. From a physical perspective, the relation may be viewed as encoding how the balance between thermal AdS and black-hole-dominated phases reorganizes under dimensional reduction. For pure Lovelock theories, the deviation from unity captures the increasing influence of higher-curvature interactions on the effective gravitational dynamics. In the joint large-$n$, large-$d$ limit, provided that $d-2n>1$, the correction factor tends to unity, thereby asymptotically recovering the Einstein-gravity relation. This suggests that sufficiently high-order pure Lovelock theories reproduce, in an asymptotic thermodynamic sense, the Einstein-gravity duality relation between the characteristic temperatures.

The inclusion of electric charge qualitatively modifies the thermodynamic structure. In the grand canonical ensemble of Einstein gravity black holes, the HP phase transition temperature and the minimum temperature continue to exhibit a remarkably simple relation: they differ only by a multiplicative factor determined solely by the spacetime dimension. More precisely, the minimum temperature in $(d+1)$ dimensions can be mapped to the Hawking--Page transition temperature in $d$ dimensions provided the fixed electrostatic potential is simultaneously rescaled by an appropriate dimension-dependent factor. Thus, even in the presence of charge, Einstein gravity retains a nontrivial remnant of the neutral-sector universality. This structure, however, does not extend to higher-curvature Lovelock theories. Once higher-curvature interactions are included, no universal, only dimension-dependent mapping between the two temperatures survives. The failure of such a correspondence reflects the nontrivial interplay between electric charge and pure Lovelock order, which introduces additional theory-dependent scales into the thermodynamic system. Consequently, the phase structure becomes substantially richer, with the stability properties depending sensitively on both the Lovelock order and the electrostatic sector. From a physical perspective, the origin of this behavior can be traced to the competing effects of electrostatic repulsion and higher-curvature contributions to the gravitational dynamics. The electric field tends to stabilize smaller black holes by counteracting gravitational attraction, thereby shifting both the onset of local thermodynamic stability and the HP transition point. In Einstein gravity, these effects preserve an underlying scaling structure, leading to the dimensionally related temperatures described above. In contrast, higher-curvature terms modify the effective gravitational interaction in a non-universal manner, obstructing any simple scaling relation and yielding a considerably more intricate thermodynamic landscape. 

The distinction from the previously known results can be seen at the level of the two defining conditions of the duality. The minimum temperature is fixed locally by the turning-point condition $(\partial T/\partial r_+)_{\Phi,P}=0$, whereas the Hawking--Page temperature is fixed globally by the vanishing of Gibbs free energy. In Einstein gravity, the radial dependence of the gravitational, AdS, and electromagnetic contributions to these two conditions is sufficiently compatible that the resulting expressions can be mapped into one another by changing the spacetime dimension together with a dimension-dependent rescaling of the electrostatic potential, as in Eq.~\eqref{eq:noveldualGRCharged}. For pure Lovelock gravity with $n>1$, this compatibility is lost: the Lovelock term changes the powers of $r_+$ entering the temperature and free energy, while the electromagnetic sector contributes with powers that combine differently in the turning-point and $G=0$ conditions. In particular, the HP condition contains a term proportional to $r_+^{2n-2}$ in addition to the $r_+^{2n}$ contribution, as made explicit in Appendix~\ref{Appen:smallpotentApprox}. Thus, changing $d$ and rescaling $\Phi$ does not simultaneously reproduce the radial structure of the two conditions. The absence of a universal charged duality for $n>1$ is therefore not merely the absence of a relation that has yet to be identified; it is a consequence of the incompatible scaling structures generated by the simultaneous presence of higher-curvature and electromagnetic contributions.

In the neutral sector, the absence of the electromagnetic scale allows the pressure dependence of $T_{\min}$ and $T_{\rm HP}$ to factorize in the same way, leaving a ratio controlled only by $d$ and $n$, as in Eq.~\eqref{eq:novelDualrelationneutral}. In Einstein gravity with fixed potential, the additional electromagnetic contribution preserves enough of this factorization to be absorbed into a dimension-dependent redefinition of $\Phi$. For $n>1$, however, the charge-dependent terms remain entangled with the Lovelock-dependent horizon-radius scaling. Consequently, the ratio $T_{\rm HP}/T_{\min}$ retains dependence on the continuous thermodynamic parameters $P$ and $\Phi$, rather than reducing to a function of $d$ and $n$ alone. 

Further insight into the phase structure can be obtained from the framework of thermodynamic information geometry, in which the Ruppeiner scalar curvature furnishes an invariant characterization of the effective statistical interactions underlying the black hole microstructures. In particular, divergences, sign changes, and extrema of the scalar curvature are known to encode critical phenomena and phase reorganizations in a manner analogous to conventional many-body systems. Evaluating the normalized Ruppeiner scalar curvature along the HP coexistence curve reveals a remarkably rigid and universal structure. For electrically neutral black holes, we find that the scalar curvature evaluated precisely at the HP transition assumes a constant value fixed entirely by the spacetime dimension and the Lovelock order, remaining completely independent of all other thermodynamic parameters such as the temperature, pressure, horizon radius, or coupling scales. This demonstrates that the HP transition is not merely a thermodynamic coexistence condition but corresponds to a geometrically distinguished hypersurface in the equilibrium state manifold, characterized by an intrinsic curvature scale determined solely by the underlying gravitational dynamics. From a microscopic perspective, this universality suggests that the collective statistical interactions governing the confined/deconfined transition are organized by the geometric structure of the gravitational theory itself rather than by state-dependent thermodynamic data. In particular, the independence from continuous thermodynamic parameters indicates that the effective correlation volume at the transition is controlled entirely by the fundamental higher-curvature theory. Such behavior strongly parallels universality at ordinary phase transition points, where macroscopic observables lose sensitivity to microscopic details and become governed only by global structural data such as symmetry and dimensionality.

Remarkably, the universality becomes even stronger in the Einstein limit when the system is analyzed in the grand canonical ensemble. In this case, the normalized scalar curvature evaluated at the HP transition remains universal even in the presence of electric charge, depending only on the spacetime dimension if the electrostatic potential is less than a critical value, expressed as $d - 2 > 2(d - 3)\Phi^2$. In this regime, the electric sector therefore does not deform the intrinsic thermodynamic geometry of the transition point, despite modifying the surrounding phase structure. This suggests that the HP transition in the grand canonical ensemble defines a genuinely geometric universality class whose thermodynamic curvature is protected against charge deformations. Such behavior provides further evidence that the HP transition is encoded in a deep geometric property of the equilibrium state space, potentially reflecting an underlying universality in the dual holographic description.

Such thermodynamic quantity independence of the normalized scalar curvature is, however, not generic in higher-order pure Lovelock black holes and therefore points to a special property of general relativity. In such cases, the scalar curvature at the HP phase transition depends on multiple parameters, including the pressure, electrostatic potential, spacetime dimension, and the specific Lovelock order. This dependence reflects the richer geometric structure induced by higher-order theories, where thermodynamic quantities are more sensitive to the full set of physical parameters. Nevertheless, an interesting universality emerges in the limit of large pressure or in the simultaneous large electrostatic potential and large pressure limit. In these regimes, the scalar curvature asymptotically approaches a constant value at the transition, indicating a form of emergent universality. Physically, this suggests that at high energy scales (large pressure) or in the presence of a strong electrostatic field (large potential) together with large pressure, the system effectively forgets the detailed structure of the underlying theory and flows toward a universal behavior. Furthermore, we also find that, in the limit of a large electrostatic potential, the HP transition becomes improbable for fixed finite pressure and temperature.  

Previous studies of Ruppeiner geometry in pure Lovelock gravity have mainly investigated the curvature over the thermodynamic state space and its behavior near critical points. In contrast, our analysis focuses specifically on the Hawking--Page transition surface defined by the condition $G=0$. We find that this restriction leads to a remarkable simplification in the neutral case, where all continuous thermodynamic parameter dependence cancels. In the charged case, this cancellation persists for Einstein gravity but is broken for higher-order Lovelock theories, where the normalized curvature retains dependence on the pressure and electrostatic potential. We found regimes where the scalar curvature asymptotically approaches a constant value at the transition, indicating a form of emergent universality. Thus, the novelty of our analysis lies not merely in computing the Ruppeiner curvature, but in revealing its special simplification on the Hawking--Page surface and identifying how this universality is precisely broken and restored in appropriate limits in higher-curvature theories.

The special case of static, spherically symmetric, pure Lovelock AdS black holes discussed in the manuscript points to a hierarchy of universality in black hole thermodynamics. General relativity exhibits the highest degree of universality, with simple, dimensionally governed relations, while higher-curvature theories introduce corrections that enrich the structure but often obscure simple scaling laws. The recovery of universal behavior within certain limits hints at deeper underlying principles, possibly connected to holography or renormalization group flows in the dual field theory. It would be interesting to further explore these connections, particularly in the context of extended phase space thermodynamics and their implications for dual strongly coupled systems.

\section*{Acknowledgements}
The authors thank Sumanto Chanda for the valuable discussions on the geometry of thermodynamics. S.K.’s research is supported by the Anusandhan National Research Foundation, India, under the ANRF ARG Grant (No. ANRF/ARG/2025/005821/PS).

\begin{appendices}

\section{Large pressure, fixed potential limit of scalar curvature} \label{Appen:smallpotentApprox}
The vanishing of Gibbs free energy shown in Eq.\eqref{eq:GibbsFreeExplicitChargAdS} at the Hawking-Page transition gives the condition:
\begin{align}\label{eq:vanishGibbsCondition}
    A_1 + B_1 r_+^{2n} + C_1 r_+^{2n-2}=0.
\end{align}
Here,
\begin{align}
    A_1=\frac{\Sigma_{d-2}(d-2)\hat{\alpha}_n}{16\pi(d-2n)}; \qquad 
    B_1 =\frac{\Sigma_{d-2}(1-2n)P}{(d-2n)(d-1)}; \qquad 
    C_1=\frac{\Sigma_{d-2}(d-3)(2n-3)\Phi^2}{8\pi(d-2n)}.
\end{align}
We can rewrite Eq.\eqref{eq:vanishGibbsCondition} as
\begin{align} \label{eq:tobeusedintermediate}
    r_+^{2n} = -\frac{A_1}{B_1}-\frac{C_1}{B_1}r_+^{2n-2}.
\end{align}
At $C_1=0$, 
\begin{align}
    r_{+0}^2 = \bigg( \frac{-A_1}{B_1} \bigg)^{1/n}.
\end{align}
Now, writing $r_+^2 = r_{+0}^2(1+u)$ in Eq.\eqref{eq:tobeusedintermediate} we get
\begin{align}
    &(r_{+0}^2(1+u))^n= -\frac{A_1}{B_1} -\frac{C_1}{B_1}r_{+0}^{2n-2}(1+u)^{n-1} \\
   & \implies (1+u)^n = 1-\frac{C_1}{B_1 r_{+0}^2}(1+u)^{n-1} \label{eq:middleApproxstep}.
\end{align}
Defining $\epsilon =\frac{C_1}{B_1 r_{+0}^2}$, we note that
\begin{align}
\epsilon \sim \frac{\Phi^2}{P\, r_{+0}^2}.
\end{align}
Using $r_{+0}^{2n} \sim \frac{1}{P}$, we obtain $r_{+0}^2 \sim P^{-1/n}$, and hence
\begin{align}
\epsilon \sim \Phi^2\, P^{-(1-1/n)}.
\end{align}
Thus, in the regime $\Phi^2 P^{-(1-1/n)} \ll 1$, we have $\epsilon \ll 1$. Set $y:=1+u$ and define
\begin{align}
F(y,\epsilon):=y^{\,n-1}(y+\epsilon)-1 .
\end{align}
For $\epsilon=0$, $F(y,0)=y^n-1$ so the $C_1=0$ branch corresponds to the root $y_0=1$ (i.e. $u_0=0$, see Eq.\eqref{eq:middleApproxstep}). Moreover
\begin{align}
\partial_y F(y,\epsilon)=(n-1)y^{n-2}(y+\epsilon)+y^{n-1}\;\Rightarrow\;\partial_yF(1,0)=n\neq 0.
\end{align}
Hence, by the Implicit Function Theorem, there exists $\delta>0$ and a unique smooth function
$y(\epsilon)$ for $|\epsilon|<\delta$ such that $F(y(\epsilon),\epsilon)=0$ and $y(\epsilon)\to 1$ as
$\epsilon\to 0$. Therefore $u(\epsilon)=y(\epsilon)-1\to 0$, i.e. $u$ is small for $|\epsilon|\ll 1$ on this branch.
Differentiating $F(y(\epsilon),\epsilon)=0$ gives
\begin{align}
y'(0)=-\frac{\partial_\epsilon F(1,0)}{\partial_y F(1,0)}=-\frac{1}{n}
\quad\Rightarrow\quad
u(\epsilon)=y(\epsilon)-1=-\frac{\epsilon}{n}+O(\epsilon^2) \label{eq:middlestep2appr}.
\end{align}
We may therefore expand
\begin{align}
(1+u)^n \approx 1+nu, \qquad (1+u)^{n-1} \approx 1+(n-1)u.
\end{align}
Substituting into Eq.\eqref{eq:middleApproxstep} gives
\begin{align}
1+nu = 1-\epsilon(1+(n-1)u).
\end{align}
Solving to leading order in $\epsilon$, we obtain
\begin{align}
u \approx -\frac{\epsilon}{n},
\end{align}
which is same as Eq.\eqref{eq:middlestep2appr}. Substituting into $r_+^2 = r_{+0}^2(1+u)$ and expanding the square root, we get
\begin{align}
r_+ = r_{+0}\sqrt{1+u} \approx r_{+0}\left(1+\frac{u}{2}\right)
= r_{+0}\left[1-\frac{1}{2n}\frac{C_1}{B_1 r_{+0}^2}\right].
\end{align}
Thus,
\begin{align}
    r_+ \approx \bigg(-\frac{A_1}{B_1}\bigg)^{1/2n} 
    \bigg[ 1-\frac{C_1}{2nB_1} \bigg(-\frac{A_1}{B_1}\bigg)^{-1/n} \bigg].
\end{align}
Substituting $A_1,B_1$ and $C_1$, we see that the second term in the parenthesis vanishes in the limit $P\rightarrow\infty$, and we have
\begin{align} \label{eq:limitlargePrplus}
    r_+ \rightarrow \bigg(-\frac{A_1}{B_1}\bigg)^{1/2n}.
\end{align}
which is the same as that for the chargeless black holes. Taking  the same limit in 
\begin{align}  \label{eq:chargedRuppienerA}
R_N=\frac{1}{2}
- \frac{1}{2
\left(
1 + \frac{(d - 2n - 1)\, r_+^{-1}}{2\pi (1 - 2n)\, T}
- \frac{(d - 3)^2\,\, \Phi^2\, r_+^{2n - 3}}{n \pi \hat{\alpha}_n (d - 2)(1 - 2n)\, T}
\right)^{1/2}
}.
\end{align}
and substituting Eq.\eqref{eq:limitlargePrplus} one gets
\begin{align} \label{eq:chargeLessRuppiener}
R_N= -\frac{(d-2n-1)(d-2n+1)}{2},
\end{align}
at the Hawking Page transition. Therefore, we conclude that in the $\Phi^2 P^{-(1-1/n)} \ll 1$ limit, the scalar curvature is a constant of the HP transition.

\section{Joint limit $\Phi\to\infty$ and $P\to\infty$ with $\Phi^2/P$ fixed} \label{Append:bothLargeLimit}

We now consider the algebraic equation for the horizon radius \( r_+ \), which follows from the condition that the Gibbs free energy vanishes at the Hawking Page transition (Eq.~\eqref{eq:GibbsFreeExplicitChargAdS}), given by:
\begin{align} 
0
&=
\frac{\Sigma_{d-2}(d-2)r_+^{d-1}}{16\pi }
\left[
\frac{\hat{\alpha}_n}{r_+^{2n}} 
+ \frac{16\pi P}{(d-1)(d-2)}
+ \frac{2(d-3) \Phi^2}{(d-2)r_+^2}
\right]
-\left[
\frac{d-2n-1}{4\pi n r_+}
+ \frac{4P}{ n \hat{\alpha}_n (d-2)}r_+^{2n-1}
- \frac{(d-3)^2  \Phi^2}{2\pi (d-2)n \hat{\alpha}_n}r_+^{2n-3}
\right]  \nonumber \\
&\qquad\qquad\qquad
\times \frac{(d-2)\Sigma_{d-2}\hat{\alpha}_n n r_+^{d-2n}}{4(d-2n)}
- \frac{\Phi^2 (d-3)\Sigma_{d-2}}{4\pi} r_+^{d-3},
\label{eq:master}
\end{align}
where $\Sigma_{d-2}$ is the volume of the unit $(d-2)$-dimensional space, $P$ is the pressure, $\Phi$ is the electric potential and $\hat\alpha_n$ is the (rescaled) Lovelock coupling. 

A physically relevant way to implement the joint limit is to send $\Phi\to\infty$ and $P\to\infty$ while keeping
\begin{equation}
\lambda \equiv \frac{\Phi^2}{P}
\end{equation}
finite. We therefore adopt the general scaling ansatz
\begin{equation}
r_+^2 \sim \lambda^{\gamma},
\qquad (\gamma\ \text{a priori undetermined}).
\label{eq:r_general_ansatz}
\end{equation}
In the reduced large-$(\Phi,P)$ equation, the leading contributions are of the form
$P\,r_+^{d-1}$ and $\Phi^2 r_+^{d-3}$. Using $\Phi^2=\lambda P$ and
$r_+^{d-1}=(r_+^2)^{\frac{d-1}{2}}$, $r_+^{d-3}=(r_+^2)^{\frac{d-3}{2}}$, we find the scalings
\begin{align}
P\,r_+^{d-1} &\sim P\,\lambda^{\gamma\frac{(d-1)}{2}},\\
\Phi^2\,r_+^{d-3} &\sim (\lambda P)\,\lambda^{\gamma\frac{(d-3)}{2}}
= P\,\lambda^{1+\gamma\frac{(d-3)}{2}}.
\end{align}
For these two leading terms to compete at the same order in the double-scaling limit, their $\lambda$-powers must match:
\begin{equation}
\gamma\frac{(d-1)}{2}=1+\gamma\frac{(d-3)}{2}
\qquad\Longrightarrow\qquad
\gamma=1.
\end{equation}
Thus, we can consider
\begin{equation}
r_+^2 \sim \lambda,
\qquad\text{i.e.}\qquad
r_+ = C\,\sqrt{\lambda}=C\,\frac{\Phi}{\sqrt{P}}.
\label{eq:r_ansatz}
\end{equation}

Keeping only the leading $P$ and $\Phi^2$ contributions in Eq.~\eqref{eq:master} (i.e.\ dropping terms that are $\mathcal{O}(1)$ in the joint limit compared to the $\mathcal{O}(P)$ pieces, since we consider $r_+^2 \sim \lambda$) yields
\begin{align}
0 &\simeq
\Sigma_{d-2}P\,r_+^{d-1}\left(\frac{1}{d-1}-\frac{1}{d-2n}\right)
+\Sigma_{d-2}\frac{(d-3)\Phi^2}{8\pi}r_+^{d-3}
\left(\frac{d-3}{d-2n}-1\right).
\label{eq:reduced_joint}
\end{align}

Since $d\neq 2n$ and $d>3$, dividing \eqref{eq:reduced_joint} by $\Sigma_{d-2}r_+^{d-3}$ gives a linear equation for $r_+^2$:
\begin{equation}
P\,r_+^2\left(\frac{1}{d-1}-\frac{1}{d-2n}\right)
=
\frac{(d-3)\Phi^2}{8\pi}\left(1-\frac{d-3}{d-2n}\right).
\label{eq:linear_r2_wrongsign_fixed_before_solve}
\end{equation}
The right-hand side of the above equation can be written as
\begin{equation}
\frac{(d-3)\Phi^2}{8\pi}\left(1-\frac{d-3}{d-2n}\right)
=
\frac{(d-3)\Phi^2}{8\pi}\left(\frac{2n-3}{2n-d}\right)
=
-\frac{(d-3)(2n-3)}{8\pi(d-2n)}\Phi^2,
\end{equation}
so that the linear equation is more transparently expressed as
\begin{equation}
P\,r_+^2\left(\frac{1}{d-1}-\frac{1}{d-2n}\right)
=
-\frac{(d-3)(2n-3)}{8\pi(d-2n)}\,\Phi^2.
\label{eq:linear_r2_correct}
\end{equation}

\paragraph{Solution for $n\ge 2$.}
For $n\ge 2$ we have $2n-3>0$ and, in the regime of interest (AdS pressure), $P>0$ and 
\begin{align}
\left(\frac{1}{d-1}-\frac{1}{d-2n}\right)=\frac{1-2n}{(d-1)(d-2n)}<0.
\end{align}
Therefore Eq.~\eqref{eq:linear_r2_correct} admits a unique positive solution for $r_+^2$:
\begin{equation}
r_+^2 \simeq \frac{\Phi^2}{P}\;
\frac{(d-3)}{8\pi}\;
\frac{(2n-3)(d-1)}{2n-1}.
\label{eq:r2_solution}
\end{equation}
Equivalently,
\begin{equation}
r_+ \simeq 
\sqrt{\frac{(d-3)(d-1)(2n-3)}{8\pi(2n-1)}}\;
\frac{\Phi}{\sqrt{P}}, \qquad \implies \qquad \,  C^2 =
\frac{(d-3)(d-1)(2n-3)}{8\pi(2n-1)},
\label{eq:r_final_joint}
\end{equation}

Plugging the above expression of outer horizon radius into the expression of scalar curvature, Eq.\eqref{eq:chargedRuppiener}, we see that, in $P\rightarrow \infty, \Phi \rightarrow \infty$ limit, $R_N$ depends only on dimensions and the theory considered. It is given by
\begin{align}
    R_N(P\rightarrow \infty,\Phi\rightarrow \infty) \approx  \frac{1}{2} - \frac{(2n-1)^2(8\pi C^2-(d-3)^2)^2}{2((2n-1)(8\pi C^2-(d-3)^2)+2(d-3)^2)^2}
\end{align}

\paragraph{Special case $n=1$.}
For $n=1$ the structure of the reduced equation changes qualitatively. Working directly with the exact large-$(\Phi,P)$ reduction of Eq.~\eqref{eq:master} (before dropping $\mathcal{O}(1)$ pieces) one finds
\begin{equation}
0=
\frac{(d-2)}{16\pi}
-\frac{1}{d-1}P\,r_+^{2}
-\frac{(d-3)}{8\pi}\Phi^2,
\end{equation}
so that
\begin{equation}
r_+^2
=
\frac{d-1}{P}\left(\frac{(d-2)}{16\pi}-\frac{(d-3)}{8\pi}\Phi^2\right).
\label{eq:r2_n1_exact}
\end{equation}
In the joint limit $\Phi^2=\lambda P$ with $P\to\infty$ at fixed $\lambda>0$, this behaves as
\begin{equation}
r_+^2
=
(d-1)\left(\frac{(d-2)}{16\pi P}-\frac{(d-3)}{8\pi}\lambda\right)
\;\xrightarrow[P\to\infty]{}\;
-(d-1)\frac{(d-3)}{8\pi}\lambda \,<\,0,
\end{equation}
and hence there is \emph{no} real positive horizon radius in this double-scaling limit for generic fixed $\lambda>0$.
Equivalently, for $n=1$ a necessary condition for a physical solution at fixed $P$ is
\begin{equation}
r_+^2>0
\qquad\Longleftrightarrow\qquad
\Phi^2<\frac{d-2}{2(d-3)}\,,
\end{equation}
which is incompatible with $\Phi\to\infty$.
Therefore the double-scaling joint limit with finite $\lambda$ is physically sensible for $n\ge 2$,
but does not yield a physical (real, positive) $r_+$ branch for $n=1$.

\section{Scalar curvature for general relativity} \label{AppendicRuppienerGRexplicit}
In general relativity ($n = 1$) the scalar curvature at the Hawking-Page transition temperature can be written from Eq.~\eqref{eq:chargedRuppiener} as
\begin{align} \label{eq:RuppiennerScalarGRa}
    R_N = \frac{1}{2} - \frac{1}{2\bigg( 1- \frac{(d-3)}{d-2-2(d-3)\Phi^2} + \frac{2(d-3)^2\Phi^2}{ (d-2)(d-2-2(d-3)\Phi^2)}  \bigg)^2},
\end{align}
Let us simplify the expression inside the brackets. Define $A = d-2 - 2(d-3)\Phi^2$ then the bracket becomes
\begin{align} \label{eq:Riccimiddlestep}
1 - \frac{d-3}{A} + \frac{2(d-3)^2\Phi^2}{(d-2)A}.
\end{align}
Taking a common denominator $(d-2)A$, and substituting $A = d-2 - 2(d-3)\Phi^2$, the numerator becomes
\begin{align}
&(d-2)(d-2 - 2(d-3)\Phi^2) - (d-3)(d-2) + 2(d-3)^2 \Phi^2\\
&=(d-2)^2 - 2(d-2)(d-3)\Phi^2 - (d-3)(d-2) + 2(d-3)^2 \Phi^2 \\
&=\big[(d-2)^2 - (d-3)(d-2)\big] + \big[-2(d-2)(d-3) + 2(d-3)^2\big]\Phi^2\\
&=(d-2) - 2(d-3)\Phi^2 = A.
\end{align}
Hence, the entire expression, Eq.\eqref{eq:Riccimiddlestep}, reduces to
\begin{align}
\frac{A}{(d-2)A} = \frac{1}{d-2}.
\end{align}
Substituting back in Eq.\eqref{eq:RuppiennerScalarGRa}, we obtain
\begin{align} \label{eq:RuppiennerScalarGRb}
R_N &= \frac{1}{2} - \frac{(d-2)^2}{2}
= -\frac{(d-3)(d-1)}{2}.
\end{align}

\end{appendices}

\bibliography{reference}

\end{document}